\documentclass{basi}
\usepackage{epsfig}
\usepackage{float}
\usepackage{amsmath}
\usepackage{rotating}
\begin{document}
\title[ ]{Further evidence for intra-night optical variability of radio-quiet 
          quasars \\} 
\author[A. Goyal]%
       {A. Goyal,$^1$\thanks{e-mail:arti@aries.ernet.in}
       Gopal-Krishna,$^2$\thanks{e-mail:krishna@ncra.tifr.res.in}
       Ram Sagar,$^1$\thanks{e-mail:sagar@aries.ernet.in}
       G.C. Anupama$^3$\thanks{e-mail:gca@iiap.res.in}
       D.K. Sahu$^3$\thanks{e-mail:dks@crest.ernet.in} \\
       $^1$ Aryabhatta Research Institute of Observational Sciences 
       (ARIES) Naini Tal 263 129 \\
       $^2$ National Centre for Radio Astrophysics (NCRA), TIFR, Pune 411 007\\
       $^3$ Indian Institute of Astrophysics (IIA) Bangalore 560 034
}
\maketitle
\label{firstpage}
\begin{abstract}
Although well established for BL Lac objects and radio-loud quasars, the 
occurrence of intra-night optical variability (INOV)
in radio-quiet quasars is still debated, primarily 
since only a handful of INOV events with good statistical significance,
albeit small amplitude, have been reported so far. This has motivated us 
to continue intra-night optical monitoring of bona-fide radio-quiet quasars
(RQQs). 
Here we present the results for a sample of 11 RQQs monitored by us on 
19 nights. On 5 of these nights a given RQQ was monitored simultaneously 
from two well separated observatories. In all, two clear cases and two probable 
case of INOV were detected. From these data, we estimate an INOV duty cycle 
of $\sim$8\% for RQQs which would increase to 19\% if the 
`probable variable' cases are also included. Such comparatively small 
INOV duty cycles for RQQs, together with the small INOV amplitudes ($\sim$1\%), 
are in accord with the previously deduced characteristics of this phenomenon.
\end{abstract}
\section{Introduction}

\hspace*{1.0cm}
The issue of the nature of the radio dichotomy of quasars continues to be debated
(e.g., Miller \& McAlister 1983; Kellermann et al. 1989; Becker et al. 1997; 
Ivezi\'{c} et al. 2002; Cirasuolo et al. 2003; Laor 2003; Barvainis et al. 
2005; White et al. 2007).  
From the distribution of QSOs over the radio-[O$_{\bf III}$] plane, 
it was found that radio-loud objects occur exclusively at high 
[O$_{\bf III}$] luminosities (Miller et al. 1993; Falcke et al. 1995; 
Xu et al. 1999). Moreover, from recent H-band imaging 
of 15 intermediate redshift RQQs, using the Hubble Space Telescope (HST), 
their host galaxies are found to be typically 0.5-1 magnitude less 
luminous compared to the hosts of radio-loud quasars (RLQs) at similar 
redshifts (Hyv\" onen et al. 2007). This, in turn, might 
suggest a difference between the masses of their central black holes 
(see also Sikora et al. 2007).  Preliminary evidence for difference 
in the central brightness profiles of the host galaxies of RLQs and RQQs 
has also been claimed (Capetti \& Balmaverde 2007). Very recently, using a 
large database from the Sloan Digital Sky Survey (SDSS; Schneider et al. 2005) 
for an optically selected quasar sample with median $z = 1.47$ and a 
luminosity range of $-$22 $<$ $M_{i}$ $<$ $-$30, Jiang et al. (2007) have 
shown that the radio-loud fraction of quasars increases with luminosity, 
but drops rapidly with redshift. Whilst all these observational trends are
 suggestive, their precise role in the core issue of the quasar radio 
dichotomy remains unclear.

\hspace*{1.0cm}
According to a widely accepted scenario, powerful jets of relativistic 
particles are accelerated by the central engines of the radio-loud quasars
(e.g. Begelman et al. 1984; Antonucci 1993; Urry \& Padovani 1995). The 
detection of well resolved faint, extended radio structure associated with
a number 
of radio-quiet quasars (Kellermann et al. 1994) indicates that the nuclei 
of RQQs, too, are probably capable of ejecting relativistic, albeit less 
powerful, jets. Supporting evidence comes from the detection of a 
relativistic jet in the VLBI observations of the RQQ PG 1407+263 (Blundell 
et al. 2003). Since INOV is now established 
as a common characteristic of jet-dominated AGN, such as BL Lacs (Miller 
et al. 1989; Jang \& Miller 1997), concerted efforts have been made to 
look for a similar signature of relativistic jet in RQQs as well (e.g.,
Gopal-Krishna et al. 1993, 1995, 2000, 2003; Sagar et al. 2004; Stalin et al. 
2004a, 2004b, 2005; Anupama \& Chokshi 1998; Jang \& Miller 1995, 1997; 
Romero et al. 1999, 2002; de Diego et al. 1998; Gupta \& Joshi 2005; 
Carini et al. 2007). The picture emerging from these studies is that, 
compared to BL Lac objects, the INOV displayed by RQQs is much more modest, 
both in amplitude and duty cycle (e.g., Stalin et al. 2005;
Carini et al. 2007). Also, there is a tendency for the INOV to be 
found in relatively nearby ($z < 1$) RQQs (e.g., Stalin et al. 2004a, b). 
Conceivably, this could be because the optical light received from distant 
($z\sim$2) quasars is strongly contaminated in the rest-frame by the 
thermal big-blue bump (Bachev et al. 2005).

The cause of the afore-mentioned marked difference between the INOV of 
BL Lacs and RQQs remains to be understood. Although the weaker and rarer 
INOV displayed by RQQs can be understood in terms of a modest misalignment 
of their jets from the line-of-sight (Gopal-Krishna et al. 2003; Stalin et al. 
2004a), the consistently small amplitude of the INOV exhibited by RQQs would 
also be in accord with its being associated with flaring hot spots on the 
accretion disk, as originally suggested by Wiita et al. (1996) (see also 
Mangalam \& Wiita 1993). 
In order to distinguish between the two competing scenarios for INOV, it is 
worthwhile to improve the statistics on the time scale, amplitude and duty 
cycle of INOV of RQQs. The present study represents our ongoing effort in that
direction, with the added novelty that a part of the RQQ monitoring 
reported here was carried out {\it simultaneously} from two well separated 
observatories.\\

\section{Observations}
\vspace{-0.5cm}
\subsection{The source sample and the instruments used}
The 11 optically bright RQQs monitored under this program were selected from 
the catalogue of V\'{e}ron-Cetty \& V\'{e}ron (2001) following the usual criteria of 
{\it K}-corrected ratio of 5 GHz to 2500 \AA~fluxes, {\it R}$^{*}$ $<$ 10.
The RQQs selected are bright enough ($m_B$ $<$ 17 mag) to attain an INOV 
detection threshold of $\sim$1-2\% using a 1-2 meter class telescope 
equipped with a CCD detector used as an N-star photometer. Also, to minimize 
contamination due to the host galaxy (Cellone et al. 2000), we limited our 
sample to intrinsically luminous AGNs with $M_{B}$ $<$ $-$23.5 mag (see also
Stalin et al. 2004a). Further, the sources lie at sufficiently high 
declinations, ensuring good visibility from the northern India. Tables 1 
and 2 list the basic data for our sample.  
As described below, for 5 of these RQQs
it was possible to carry out intra-night monitoring simultaneously from
two observatories, in an attempt to minimize the possibility of spurious
detection of low-amplitude variability, owing to factors like large
intra-night seeing variations, passing clouds, or some unknown instrumental
problems. Table 1 lists the basic data for the RQQs.

\begin{table}[h] \caption{The sample of 11 optically luminous radio-quiet
quasars monitored in the present study.}
\begin{tabular}{p{1.3cm}p{0.6cm}ccrp{1.3cm}p{0.5cm}ccr}\\
\hline
IAU Name & {\it B}&  {$M_{B}$} & z &  log $R^{*\dag}$& IAU Name & {\it B}&
{$M_{B}$} & z &   log $R^{*\dag}$ \\
\hline
0043$+$039 &16.00&$-$26.0&0.385&$-$0.7 &  1116$+$215 &14.59&$-$25.3&0.177&$-$0.1 \\
0100$+$020 &15.97&$-$25.2&0.393&$<$$-$0.2 & 1244$+$026 &18.30&$-$25.8&0.934&$<$0.1\\
0236$-$002 &15.70&$-$25.4&0.261&$<$$-$0.8 & 1526$+$285 &16.34&$-$25.8&0.450&$-$0.2 \\
0514$-$005 &15.87&$-$25.1&0.291&$<$$-$0.7& 1629$+$296 &17.08&$-$24.2&0.256&$<$$-$0.2\\
0748$+$295 &15.97&$-$28.4&0.914&$<$$-$0.4 & 1630$+$377 &16.62&$-$28.6&1.478&$<$$-$0.6\\
0824$+$098 &15.50&$-$25.6&0.260&0.5\\
\hline
\end{tabular}
{$\dag${\it R$^{*}$} is the {\it K}-corrected ratio of the 5 GHz to 2500
\AA~flux densities (Stocke et al. 1992)} $;$ references for the radio
fluxes are V\'{e}ron-Cetty \& V\'{e}ron (2001), NVSS (Condon et al. 1998)
and FIRST (Becker et al. 1995$;$ Bauer et al. 2000) and the
references therein. The absolute magnitude $M_{B}$ is computed taking
H$_0$ $=$ 50 kms$^{-1}$Mpc$^{-1}$, q$_0$ $=$ 0, and an optical spectral index
$\alpha$ $=$ 0.3 (defined as S $\propto$ $\nu^{-\alpha}$; Francis et al. 1991).
\end{table}

Part of the observations were carried out using 104-cm 
Sampurnanand telescope (ST) located at
Aryabhatta Research Institute of observational sciencES (ARIES), 
Naini Tal, India. It has a Ritchey-Chretien (RC) optics with 
an f$/$13 beam (Sagar 1999). The detector used was a cryogenically 
cooled 2048 $\times$ 2048 chip mounted at the Cassegrain focus. 
This chip has a readout noise of 5.3 e$^{-}$/pixel and a gain of 
10 e$^{-}$$/$Analog to Digital Unit (ADU) in the usually employed slow 
readout mode. Each pixel has a dimension of 24 $\mu$m$^{2}$ which corresponds 
to 0.37 arcsec$^{2}$ on the sky, covering a total field of 13$^{\prime}$ $\times$
13$^{\prime}$. Observations were carried out in 2 $\times$ 2 binned mode to 
improve the S$/$N ratio. All the observations were carried out using {\it R} 
filter for which the CCD sensitivity is maximum.
The seeing mostly ranged between 
$\sim$1.5$^{\prime\prime}$ to $\sim$3$^{\prime\prime}$, as determined using 3 fairly bright stars
on the CCD frame and the plots of the seeing are provided for some of the nights
in the bottom panels of Figs. 1 and 2 (see Sec. 3).

The other telescope used by us is the 201-cm Himalayan Chandra Telescope
(HCT) located at Indian Astronomical Observatory (IAO), Hanle, India which 
is also of the RC design with an f$/$9 beam at the
Cassegrain focus\footnote{http://www.iiap.res.in/$\sim$iao}. 
The detector was a cryogenically cooled 2048 $\times$ 4096 chip, of which the 
central 2048 $\times$ 2048 pixels were used. The pixel size is 15 $\mu$m$^{2}$ 
so that the image scale of 0.29 arcsec$/$pixel
covers an area of 10$^{\prime}$ $\times$ 10${^\prime}$ on the sky. The readout noise 
of CCD is 4.87 e$^{-}$/pixel and the gain is 1.22 e$^{-}$$/$ADU. The CCD 
was used in an unbinned mode. The observations were primarily made using 
{\it R} filter, except on 4 nights when  {\it V} filter was used. The
seeing ranged mostly between $\sim$1.0$^{\prime\prime}$ to $\sim$2.5$^{\prime\prime}$.\\
The exposure time was typically 12-30 minutes for the ARIES observations and 
ranged from 3-6 minutes for observations from IAO, depending on the 
brightness of the source, phase of moon and the sky transparency for 
that night. The field positioning was adjusted so as to also have within the 
CCD frame 2-3 comparison stars within about a magnitude of the RQQ, in order 
to minimize the possibility of getting spurious variability detection
(Cellone et al. 2007). For both the telescopes, the bias frames, taken 
intermittently, and twilight sky flats were obtained.

\subsection{Data reduction}
The preprocessing of images (bias subtraction, flat-fielding and cosmic-ray 
removal) was done by applying the regular procedures in IRAF\footnote{\textsc {Image Reduction 
and Analysis Facility}} and MIDAS\footnote{\textsc {Munich Image and Data Analysis System}} 
softwares. The instrumental magnitudes of the RQQ and the stars in the image frames
were determined by aperture photometry, using DAOPHOT \textrm{II}\footnote{\textsc {Dominion
Astrophysical Observatory Photometry} software} (Stetson 1987). 
The magnitude of the RQQ was measured relative to the nearly steady comparison
stars present on the same CCD frame (Table 2). This way, Differential Light 
Curves (DLCs) of each RQQ were produced relative to 2-3 comparison stars.
For each night, the selection of optimum aperture radius was done on the 
basis of the observed dispersions in the star-star DLCs for different aperture 
radii starting from the median seeing (FWHM) value on that night to 4 times 
that value.  The aperture selected was the one which showed minimum scatter 
for the steadiest DLC found for the various pairs of the comparison stars
(Stalin et al. 2004a).
\\
\begin{table}
\caption{Positions and magnitudes of the RQQs and the comparison stars used 
in the present study$^*$.}
\begin{tabular}{ccccccr}\\
\hline
IAU Name & Object & RA(J2000) &Dec(J2000) &  {\it B} & {\it R} & \it {B-R} \\
         &      &           &            & (mag)    & (mag)   & (mag) \\
\hline

0043$+$039& RQQ & 00{$^h$}45{$^m$}47{$^s$}.23&$+$04{$^\circ$}10{$^\prime$}23{$^{\prime\prime}$}.2  &16.50 & 15.59 & 0.91 \\
          & S1 &00{$^h$}45{$^m$}44{$^s$}.87  & $+$04{$^\circ$}10{$^\prime$}57{$^{\prime\prime}$}.9 &17.34&16.07 &1.27  \\
         & S2 &00{$^h$}45{$^m$}44{$^s$}.16  & $+$04{$^\circ$}13{$^\prime$}26{$^{\prime\prime}$}.0   &17.76&15.21 &2.55  \\
         & S3 &00{$^h$}45{$^m$}43{$^s$}.81  & $+$04{$^\circ$}12{$^\prime$}32{$^{\prime\prime}$}.1   &17.76&15.21 &2.55  \\
0100$+$020 & RQQ & 01{$^h$}03{$^m$}12{$^s$}.98&$+$02{$^\circ$}21{$^\prime$}10{$^{\prime\prime}$}.1  & 17.39 & 17.02 & 0.37\\
           & S1 &01{$^h$}03{$^m$}27{$^s$}.92  & $+$02{$^\circ$}24{$^\prime$}49{$^{\prime\prime}$}.4 &17.05&16.09 &0.96  \\
         & S2 &01{$^h$}03{$^m$}28{$^s$}.26  & $+$02{$^\circ$}20{$^\prime$}19{$^{\prime\prime}$}.3   &17.99&16.52 &1.47  \\
         & S3 &01{$^h$}03{$^m$}02{$^s$}.69  & $+$02{$^\circ$}18{$^\prime$}08{$^{\prime\prime}$}.1   &16.28&15.48 &0.80  \\
0236$-$001 &RQQ& 02{$^h$}39{$^m$}22{$^s$}.84&$-$00{$^\circ$}01{$^\prime$}19{$^{\prime\prime}$}.4 &15.76 & 15.07 & 0.69  \\
           &S1 &02{$^h$}39{$^m$}32{$^s$}.39  & $-$00{$^\circ$}00{$^\prime$}42{$^{\prime\prime}$}.2   &16.27&15.16 &1.11  \\
         & S2 &02{$^h$}39{$^m$}38{$^s$}.60  & $-$00{$^\circ$}04{$^\prime$}24{$^{\prime\prime}$}.0   &16.31&15.61 &0.70  \\
0514$-$005 &RQQ &05{$^h$}16{$^m$}33{$^s$}.49&$-$00{$^\circ$}27{$^\prime$}13{$^{\prime\prime}$}.5 &15.87&  15.43 &0.44  \\
           & S1 &05{$^h$}16{$^m$}27{$^s$}.61  & $-$00{$^\circ$}30{$^\prime$}54{$^{\prime\prime}$}.3 &15.25&14.52 &0.73  \\
         & S2 &05{$^h$}16{$^m$}22{$^s$}.35  & $-$00{$^\circ$}29{$^\prime$}48{$^{\prime\prime}$}.0   &16.45&15.39 &1.06  \\
         & S3 &05{$^h$}16{$^m$}21{$^s$}.11  & $-$00{$^\circ$}31{$^\prime$}04{$^{\prime\prime}$}.3   &15.48&14.64 &0.84  \\
0748$+$295 &RQQ &07{$^h$}51{$^m$}12{$^s$}.29&$+$29{$^\circ$}19{$^\prime$}38{$^{\prime\prime}$}.4 &15.97 & 15.40 & 0.57 \\
           & S1 &07{$^h$}51{$^m$}06{$^s$}.02  & $+$29{$^\circ$}16{$^\prime$}13{$^{\prime\prime}$}.4 &17.04&15.06 &1.98  \\
         & S2 &07{$^h$}51{$^m$}09{$^s$}.26  & $+$29{$^\circ$}16{$^\prime$}19{$^{\prime\prime}$}.0   &16.25&15.13 &1.12  \\
         & S3 &07{$^h$}51{$^m$}02{$^s$}.54  & $+$29{$^\circ$}19{$^\prime$}24{$^{\prime\prime}$}.0   &15.61&14.56 &1.05  \\
0824$+$098 & RQQ& 08{$^h$}27{$^m$}40{$^s$}.18&$+$09{$^\circ$}42{$^\prime$}08{$^{\prime\prime}$}.3 &16.68 & 15.32 & 1.36 \\
           & S1 &08{$^h$}27{$^m$}51{$^s$}.44  & $+$09{$^\circ$}45{$^\prime$}21{$^{\prime\prime}$}.7 &16.84&16.02 &0.82  \\
         & S2 &08{$^h$}27{$^m$}48{$^s$}.64  & $+$09{$^\circ$}45{$^\prime$}14{$^{\prime\prime}$}.4   &16.45&15.46 &0.99  \\
         & S3 &08{$^h$}27{$^m$}44{$^s$}.30  & $+$09{$^\circ$}45{$^\prime$}05{$^{\prime\prime}$}.6   &16.28&15.44 &0.84  \\
1116$+$215 &RQQ & 11{$^h$}19{$^m$}08{$^s$}.68&$+$21{$^\circ$}19{$^\prime$}18{$^{\prime\prime}$}.1 &15.36 & 14.44 &0.92  \\
           & S1 &11{$^h$}19{$^m$}19{$^s$}.08  & $+$21{$^\circ$}26{$^\prime$}20{$^{\prime\prime}$}.6 &15.42&14.21 &1.21  \\
         & S2 &11{$^h$}19{$^m$}17{$^s$}.85  & $+$21{$^\circ$}27{$^\prime$}38{$^{\prime\prime}$}.2   &16.10&14.88 &1.22  \\
         & S3 &11{$^h$}19{$^m$}24{$^s$}.62  & $+$21{$^\circ$}25{$^\prime$}08{$^{\prime\prime}$}.8   &15.54&14.35 &1.19  \\
1244$+$026 &RQQ &12{$^h$}46{$^m$}41{$^s$}.69&$+$02{$^\circ$}24{$^\prime$}11{$^{\prime\prime}$}.9 &17.65 & 17.25 & 0.40 \\
           & S1 &12{$^h$}47{$^m$}02{$^s$}.84  & $+$02{$^\circ$}25{$^\prime$}20{$^{\prime\prime}$}.2 &17.10&16.04 &1.06  \\
         & S2 &12{$^h$}46{$^m$}48{$^s$}.83  & $+$02{$^\circ$}23{$^\prime$}37{$^{\prime\prime}$}.5   &17.47&16.23 &1.24  \\
         & S3 &12{$^h$}46{$^m$}48{$^s$}.43  & $+$02{$^\circ$}20{$^\prime$}41{$^{\prime\prime}$}.7   &17.09&16.04 &1.05  \\
1526$+$285 &RQQ &15{$^h$}28{$^m$}40{$^s$}.62&$+$28{$^\circ$}25{$^\prime$}30{$^{\prime\prime}$}.0 &17.29 & 16.28 &1.01  \\
           & S1 &15{$^h$}28{$^m$}13{$^s$}.06  & $+$28{$^\circ$}27{$^\prime$}51{$^{\prime\prime}$}.8 &18.03&15.88 &2.15  \\
         & S2 &15{$^h$}28{$^m$}30{$^s$}.30  & $+$28{$^\circ$}30{$^\prime$}04{$^{\prime\prime}$}.2   &17.15&15.22 &1.93  \\
         & S3 &15{$^h$}28{$^m$}55{$^s$}.02  & $+$28{$^\circ$}25{$^\prime$}12{$^{\prime\prime}$}.1   &18.09&16.03 &2.06  \\
1629$+$299 &RQQ &16{$^h$}31{$^m$}24{$^s$}.43&$+$29{$^\circ$}53{$^\prime$}01{$^{\prime\prime}$}.7 &17.08 & 16.53 &0.55  \\
           & S1 &16{$^h$}31{$^m$}31{$^s$}.70  & $+$29{$^\circ$}50{$^\prime$}01{$^{\prime\prime}$}.2 &16.97&15.07 &1.90  \\
         & S2 &16{$^h$}31{$^m$}19{$^s$}.90  & $+$29{$^\circ$}52{$^\prime$}29{$^{\prime\prime}$}.2   &17.60&16.47 &1.13  \\
         & S3 &16{$^h$}31{$^m$}18{$^s$}.42  & $+$29{$^\circ$}52{$^\prime$}18{$^{\prime\prime}$}.8   &17.88&16.39 &1.49  \\
1630$+$377 &RQQ &16{$^h$}32{$^m$}01{$^s$}.10&$+$37{$^\circ$}37{$^\prime$}50{$^{\prime\prime}$}.1 &16.14 & 16.07 &0.07 \\
           & S1 &16{$^h$}32{$^m$}06{$^s$}.40  & $+$37{$^\circ$}39{$^\prime$}18{$^{\prime\prime}$}.1 &16.67&15.49 &1.18  \\
         & S2 &16{$^h$}32{$^m$}02{$^s$}.70  & $+$37{$^\circ$}35{$^\prime$}19{$^{\prime\prime}$}.9   &16.35&15.63 &0.72  \\
         & S3 &16{$^h$}31{$^m$}39{$^s$}.98  & $+$37{$^\circ$}36{$^\prime$}03{$^{\prime\prime}$}.9   &15.72&15.28 &0.44  \\ 
\hline
\end{tabular}

$^*$ taken from United States Naval Observatory-B catalogue
(Monet et al. 2003)
\end{table}

\section{Results}\label{sec:result}
The DLCs are shown in Figures 1 and 2 for
two station and single station monitoring, respectively, while 
tables 3 and 4 summarize the observations for the present sample of 
RQQs monitored from two stations and single station, respectively 
. For each night of observations
these tables provide the object name, date of observation,
telescope used (Table 4 only), filter(s) used, number of data points (N points),
durations of observation,
the measures of variability, C$_{eff}$ and $\psi$ and an indicator of variability status .  
The classification `variable' (V) or `non-variable' (N) was decided
using a parameter C$_{eff}$, basically defined following the criteria of 
Jang \& Miller (1997). We define C for a given DLC as the ratio of it's 
standard deviation, $\sigma$$_T$ and $\eta\sigma_{err}$, where $\sigma_{err}$
is the average of the rms errors of its individual data points and $\eta$ 
was estimated to be 1.5 (Stalin et al. 2004a, 2004b, 2005; Gopal-Krishna et al. 2003; 
Sagar et al. 2004). 
However, our analysis for the present dataset yields $\eta = 1.3$ and we
have adopted this value here. We compute C$_{eff}$ 
from the C values (as defined above) found for the DLCs of an AGN relative to
different comparison stars monitored on a given night (details are given in 
Sagar et al. 2004). This has the advantage of using multiple DLCs of
an AGN, relative to the different comparison stars. The source is termed
`V' for C$_{eff}$$ > $ 2.576, corresponding to a confidence 
level of 99\%. We call the source to be `probable variable' (PV) if C$_{eff}$ 
is in range of 1.950 to 2.576, corresponding to a confidence 
level between 95\% to 99\%. Finally, the peak-to-peak INOV amplitude is 
calculated using the definition (Romero et al. 1999) 

\begin{equation}
\psi= \sqrt{({D_{max}}-{D_{min}})^2-2\sigma^2}
\end{equation}
with \\
$D_{max}$ = maximum in the AGN differential light curve \\
$D_{min}$ = minimum in the AGN differential light curve \\
$\sigma^2$= $\eta^2$$\langle\sigma^2_{err}\rangle$
\\

Notes are given below for the RQQs showing INOV (or probable variables).

{\bf RQQ 0514$-$005$:$} This RQQ has earlier been monitored on three nights 
by Stalin et al. (2004a) who did not detect INOV. In our monitoring on 
three nights, INOV was detected on the night of 2003 Nov 20, using the HCT 
(Fig. 2), with an amplitude  $\psi \geq 1.5\%$. On another night, 2004 Nov 18,
although correlated INOV with an amplitude $\psi$ $\geq$ 1.2\% was seen in 
both QSO-star DLCs, the computed C$_{eff}$ = 1.73 falls below the formal 
qualifying threshold value of 1.950 for `PV' classification (see above).\\

{\bf RQQ 0824$+$098$:$} This source was monitored by Gopal-Krishna et al.
(2000) on one night for 3.3 hours when it showed a flare of $\sim$0.055 mag 
in just half an hour. Being a single point excursion, however, the variation 
was not considered by them as confirmed. Subsequently, the RQQ was monitored by 
Stalin et al. (2005) on one night when INOV was clearly seen, with an amplitude
$\psi \sim$2.2\%. We have monitored the source on 2005 Jan 13 using ST and HCT 
telescopes simultaneously and both sets of DLCs show a brightening by $\sim$1\% over the 
time interval 16.5-20.0 UT (Fig. 1). The sky condition remained clear at 
both the observatories and the ``seeing'' was also fairly constant, as seen 
from the bottom panels. The small brightening seen from the two observatories 
is correlated both in amplitude as well as time, corresponds to 
C$_{eff}$ = 2.66 for the HCT light curves (Table 3; Fig. 1), qualifying it as 
a positive detection of INOV. Although C$_{eff}$ = 2.12 found for the ST
data alone would qualify this source as only a `PV', 
taken together with the simultaneous HCT light curves, the ST data further 
strengthen the case for a positive detection of INOV. 

{\bf RQQ 1116$+$215$:$} This source was monitored by us on total of four nights.
On one night, observations could be made simultaneously using 
the ST and HCT (Table 3). On two nights of monitoring (i.e. on 2005 Apr 14 and 
2006 Mar 31), when it could be observed using 
ST alone, it showed 
a variation in amplitude by $\sim$1\% (Fig. 2) and is designated a `PV' on both the nights, 
since the variation is significant at confidence level of 95\% (Table 4; Fig. 2). 
On these nights, sky at the observatory was clear and the
``seeing'' was fairly constant, as seen from the bottom panels (Fig. 2). 
Unfortunately, on the night of simultaneous two-station monitoring, the sky conditions 
were not good, as the one or the other site had thin clouds throughout the 
observing run. 
\\
\begin{table}[ht]
\caption{Observation log and variability results for two station (simultaneous) monitoring.}
\begin{tabular}{cccccccc}\\
\hline
IAU Name &  Date & Filter & N points & Duration(h)& C$_{eff}$ & $\psi$\% &Status$^{\ddag}$ \\
         & yy.mm.dd&ST, HCT   & ST, HCT&ST,  HCT &ST, HCT & ST, HCT &ST, HCT \\ \hline
0100$+$020 &  05.11.05 & R, R & 20, 21 &6.0, 5.9 & 0.18, 0.38& 1.22, 1.0 & N, N \\
0236$-$002 &  05.11.06 & R, R & 13, 19 &6.4, 6.4 & 1.27, 1.80 &0.5, 0.8 & N, N \\
0748$+$295 &  04.12.17 & R, V & 17, 15 &6.8, 4.0 &1.35, 1.78 &0.5, 0.7 & N, N \\
0824$+$098 &  05.01.13 & R, V & 17, 15 &7.0, 4.0 & 2.12, 2.66 &0.9, 1.0 &PV, V \\
1116$+$215 &  06.03.07 & R, R & 46, 31 &8.3, 8.1 &1.63, 1.78 & 0.8, 0.7&N, N \\
\hline
\end{tabular}

{$^{\ddag}$V = variable; N = non-variable; PV = probable variable } 
\end{table}
\begin{table}[ht]
\caption{Observation log and variability results for single station monitoring.}
\begin{tabular}{ccccccccc}\\
\hline
IAU Name &  Date & Telescope& Filter &  N points & Duration(h) & C$_{eff}$ & $\psi$\% &Status$^{\ddag}$ \\
         & yy.mm.dd&          &        &           &             &              &           &          \\
\hline
0043$+$039 & 04.10.16 & HCT& V & 25 & 6.0 & 1.50 &0.7 & N \\
0514$-$005 & 03.11.20 & HCT& R & 39 & 7.2 & 2.70 &1.5 & V \\
           & 04.11.18 & ST & R & 18 & 3.8 & 1.73 & 1.2& N \\
           & 04.12.16 & HCT& R & 34 & 6.8 & 1.02 &1.1 & N \\
0748$+$295 & 05.01.12 & ST & R & 16 & 7.1 & 1.07 &0.5 & N \\
1116$+$215 & 05.04.14 & ST & R & 30 & 5.0 & 2.37 &1.0 & PV\\
           & 06.03.30 & ST & R & 40 & 6.2 & 1.75 &0.8 & N \\
           & 06.03.31 & ST & R & 26 & 4.2 & 2.05 &0.8 & PV\\
1244$+$026 & 05.04.13 & ST & R & 10 & 5.5 & 0.27 &0.4 & N \\
1526$+$285 & 05.05.10 & ST & R & 16 & 7.8 & 0.67 &0.5 & N \\
1629$+$299 & 04.06.15 & HCT& V & 28 & 6.1 & 1.78 &1.5 & N \\
           & 05.05.11 & ST & R & 28 & 7.7 & 0.42 &0.9 & N \\
           & 05.06.01 & ST & R & 15 & 7.4 & 1.34 &1.2 & N \\
1630$+$377 & 05.05.12 & ST & R & 29 & 6.6 & 0.31 &0.6 & N \\
\hline
\end{tabular}

{$^{\ddag}$V = variable; N = non-variable; PV = probable variable } 
\end{table}
\section{Discussion}
The duty cycle (DC) of INOV has been computed according to the procedure given 
in Romero et al. (1999)
\begin{equation}
DC  = 100\frac{\sum_{i=1}^n N_i(1/\Delta t_i)}{\sum_{i=1}^n (1/\Delta t_i)}\%
\end{equation}
where $\Delta t_i = \Delta t_{i,obs}(1+z)^{-1}$ is the
duration of monitoring session of a source on the $i$th night, corrected for 
the cosmological redshift $z$. Note that since the sources have not been 
monitored for the same duration on each night, the computation is weighted 
by the duration of monitoring $\Delta t_i$. $N_i$ was set equal to 1 if INOV 
was detected, otherwise $N_i$ = 0.
The computed duty cycle of INOV for the present sample is DC = 8$\%$. 
Note that in case 
of INOV detection from simultaneous monitoring, the data used for calculation 
is the one for which the duration is longer. Note that DC 
increases to 19$\%$
if the two cases of probable variability (i.e. RQQ 1116+215 observed 
on 2005 Apr 14 and 2006 Mar 31) are included (Table 4). These estimates are consistent with 
the previous results for RQQs (see Sec. 1). In contrast, it was found that
for BL Lac objects monitored for longer than about 4 hours, INOV with
amplitudes in excess of $3\%$ occurs with a duty cycle DC $\sim$50\% 
(e.g., Gopal-Krishna et al. 2003; Sagar et al. 2004). Thus, the observations reported here 
further strengthen the view that INOV of 
RQQs is a comparatively rare phenomenon and the variability 
amplitude remains below $3\%$. Simultaneous two-station monitoring of
such quasars would be a worthwhile effort. An added motivation
for this comes from the detection of near-infrared flares from the central 
source of our galaxy, SgrA*, on time scale of $\sim$10 minutes; these 
flares might be powered by annihilation of magnetic field near the inner 
edge of the accretion disk around the supermassive black hole in SgrA* 
(Trippe et al. 2007; Eckart et al. 2006).
\\
\section*{Acknowledgements}
We are very thankful to the anonymous referee whose valuable suggestions has
improved the quality of the paper. 
This work has made use of NASA/IPAC Extra-galactic Database (NED), which is
operated by Jet Propulsion Laboratory, California Institute of Technology, under
contract with the National Aeronautics and Space Administration.   
\newpage
\begin{figure}[h]
\hbox{
\hspace*{-15mm}{\includegraphics[height=10.0cm,width=07.0cm]{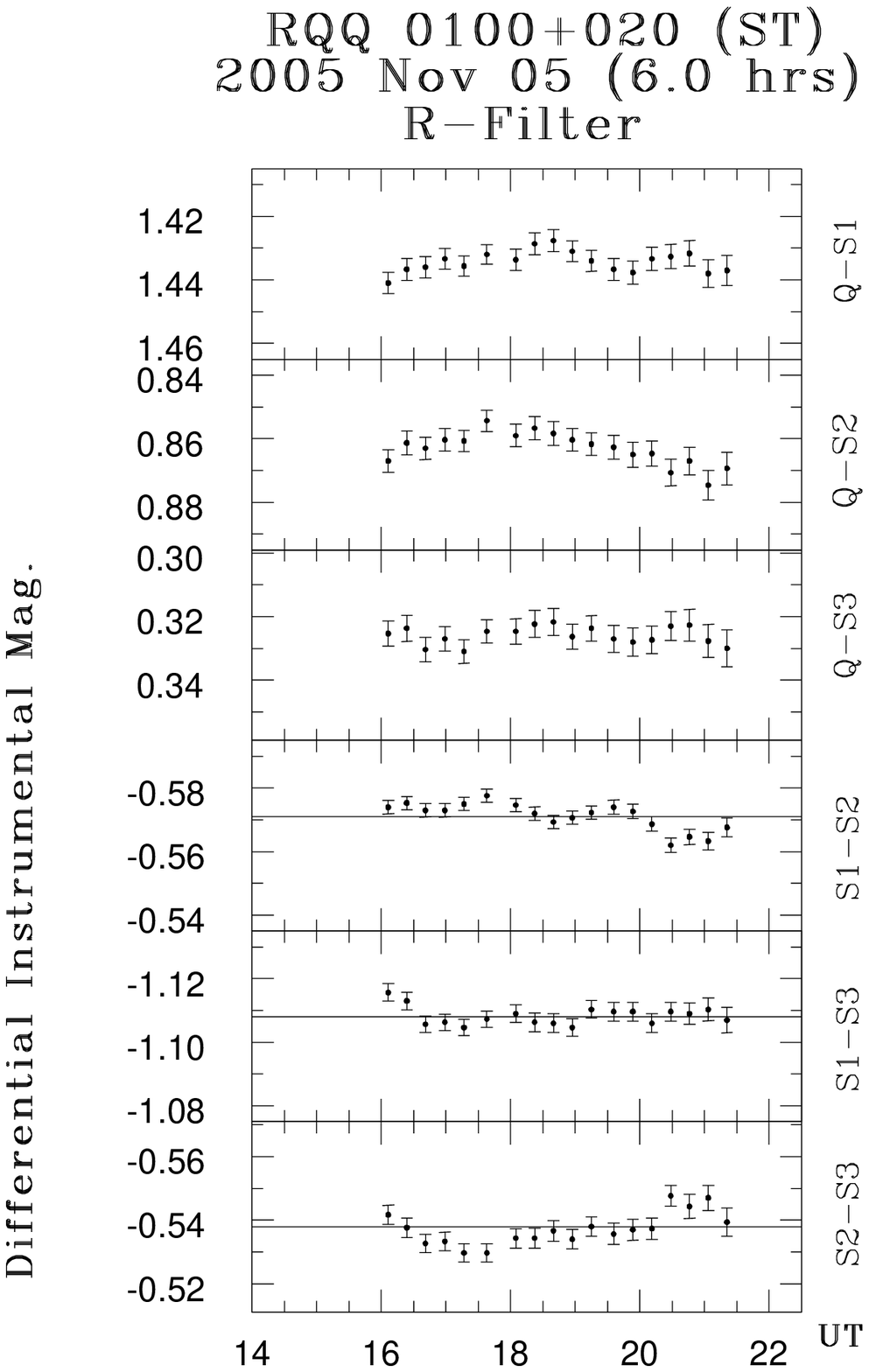}}
\hspace*{-30mm}{\includegraphics[height=10.0cm,width=07.0cm]{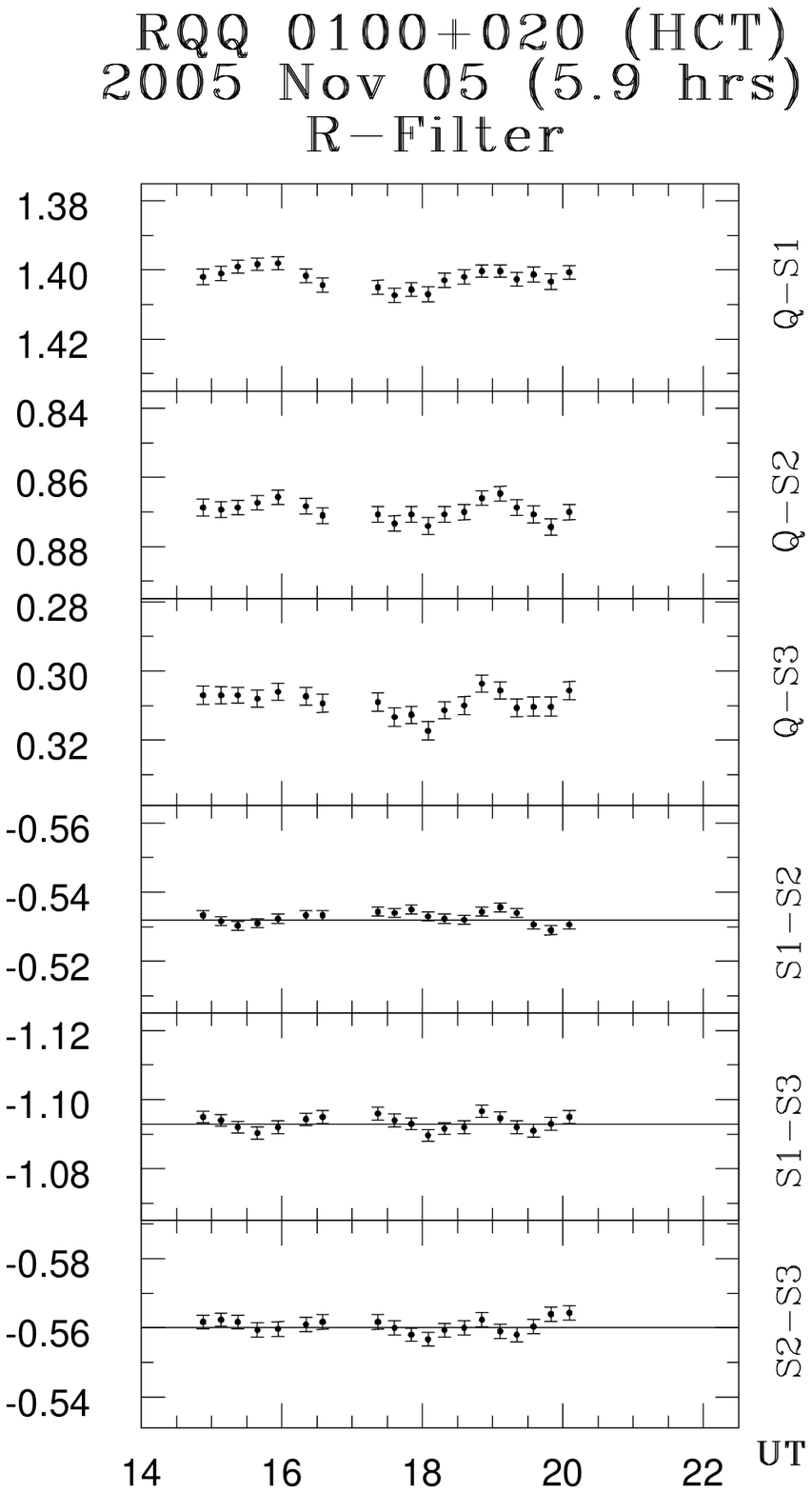}}
\hspace*{-30mm}{\includegraphics[height=10.0cm,width=07.0cm]{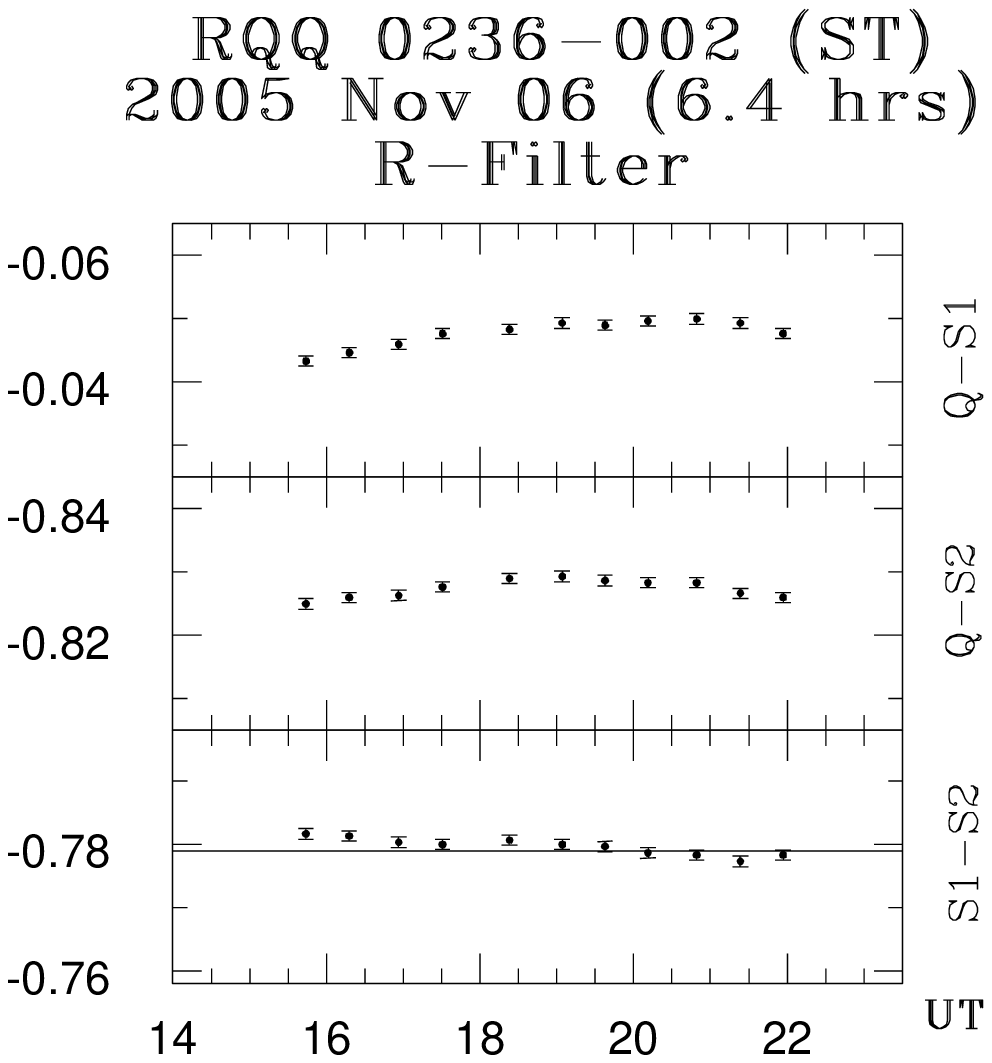}}
\hspace*{-30mm}{\includegraphics[height=10.0cm,width=07.0cm]{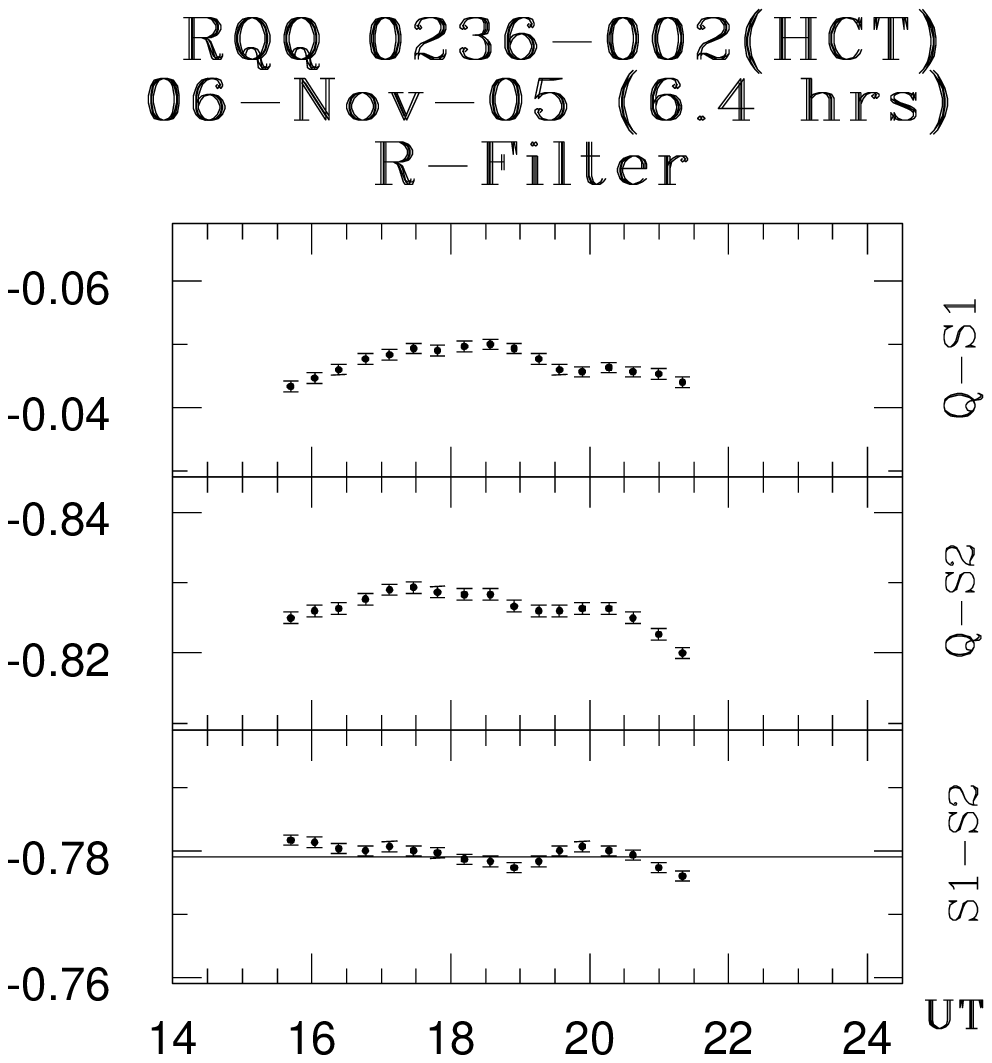}}
}
\hbox{
\hspace*{-15mm}{\includegraphics[height=10.0cm,width=07.0cm]{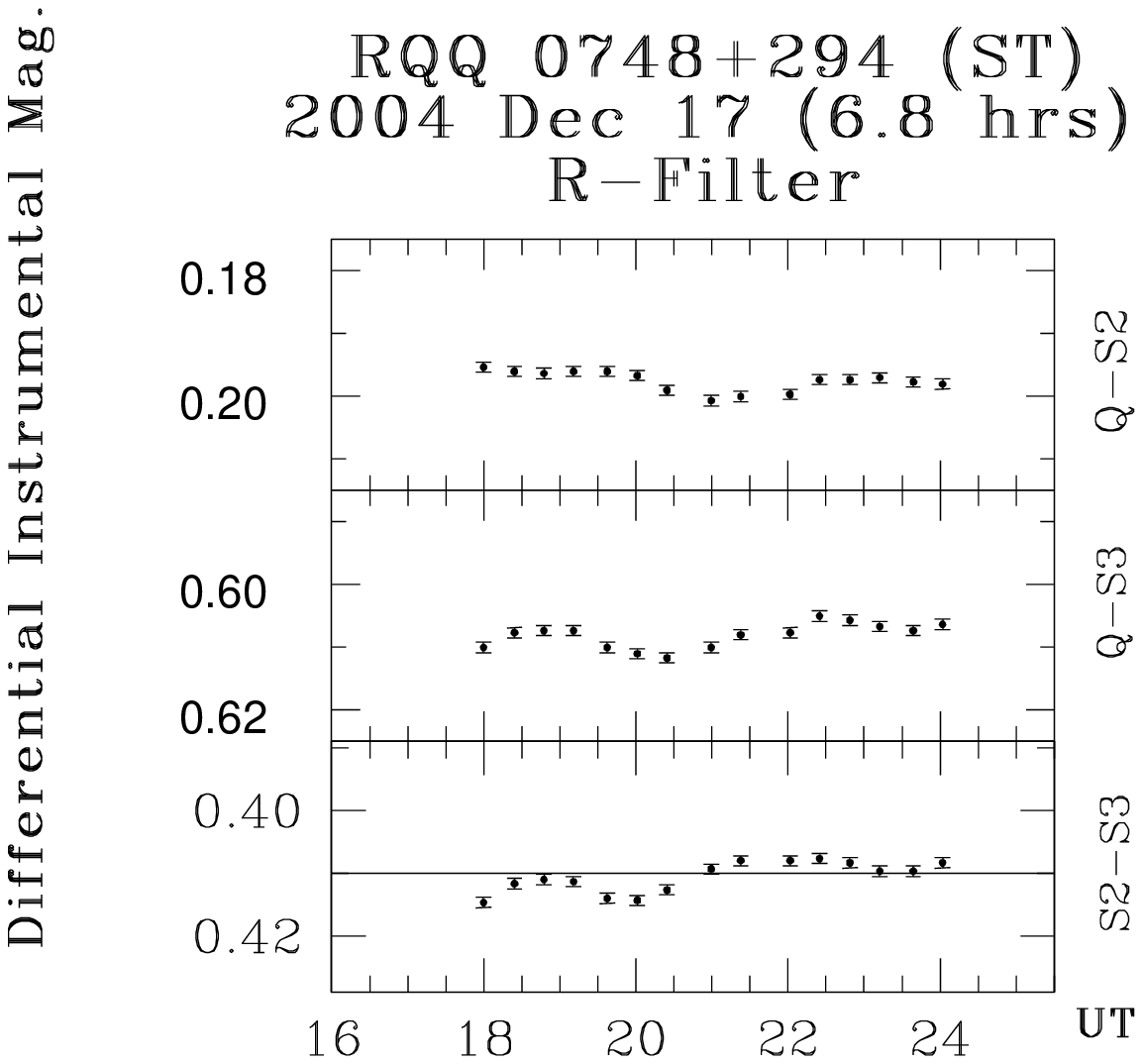}}
\hspace*{-30mm}{\includegraphics[height=10.0cm,width=07.0cm]{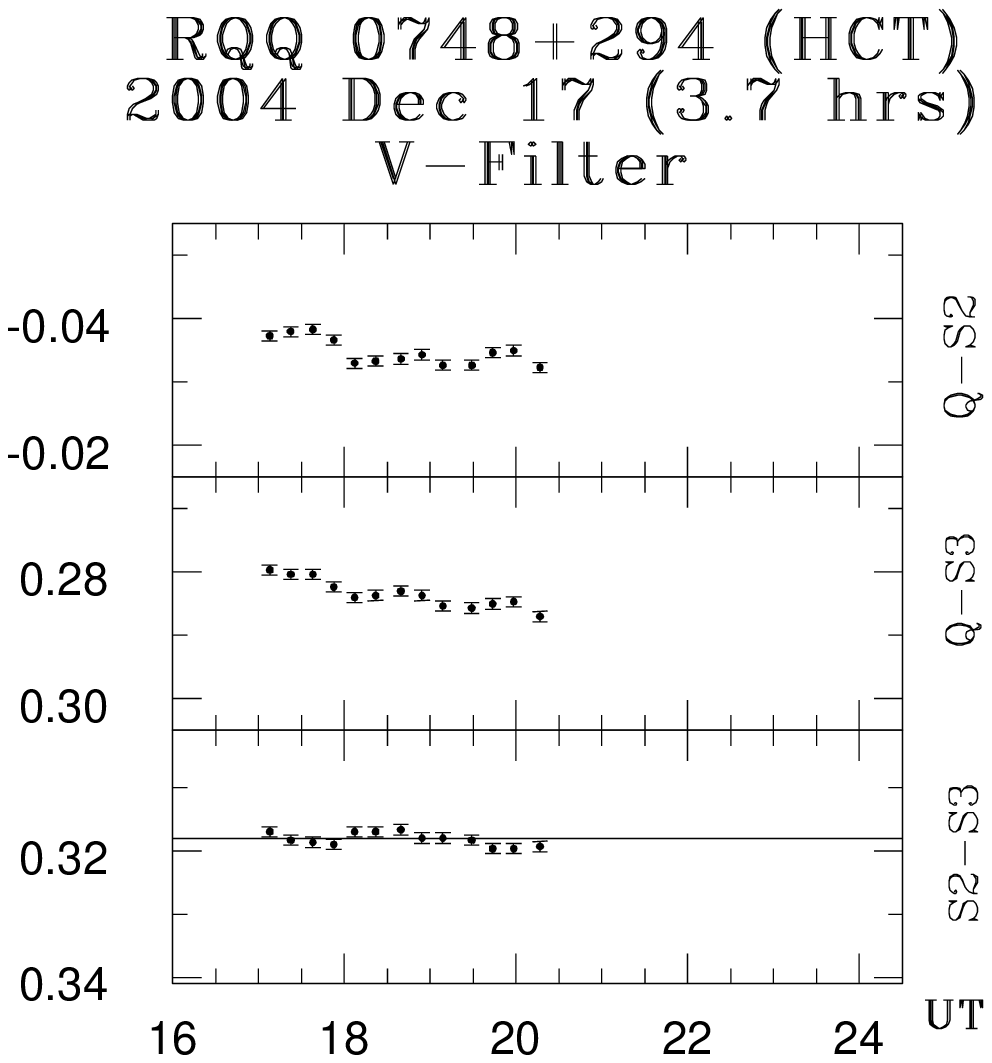}}
\hspace*{-30mm}{\includegraphics[height=10.0cm,width=07.0cm]{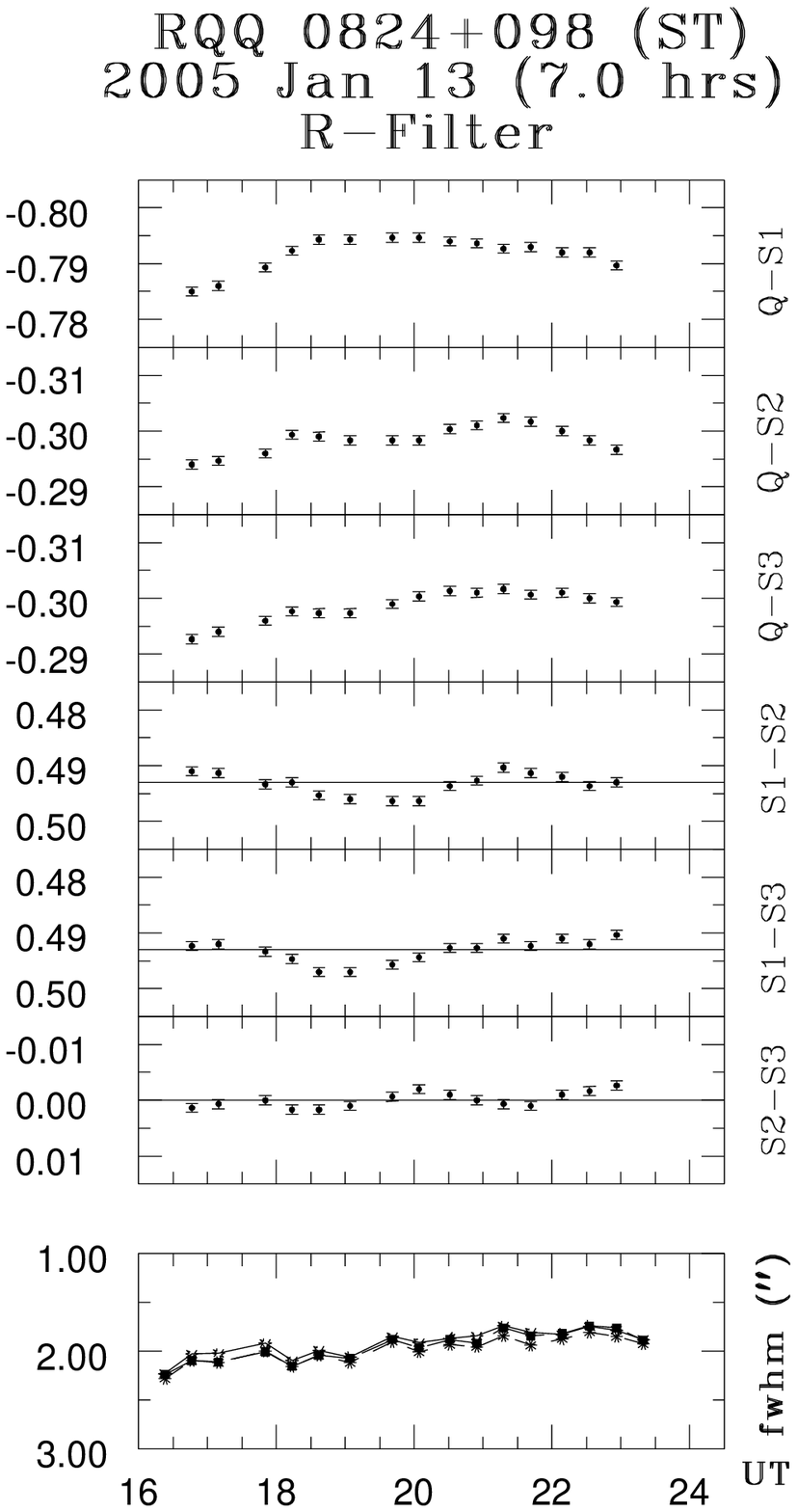}}
\hspace*{-30mm}{\includegraphics[height=10.0cm,width=07.0cm]{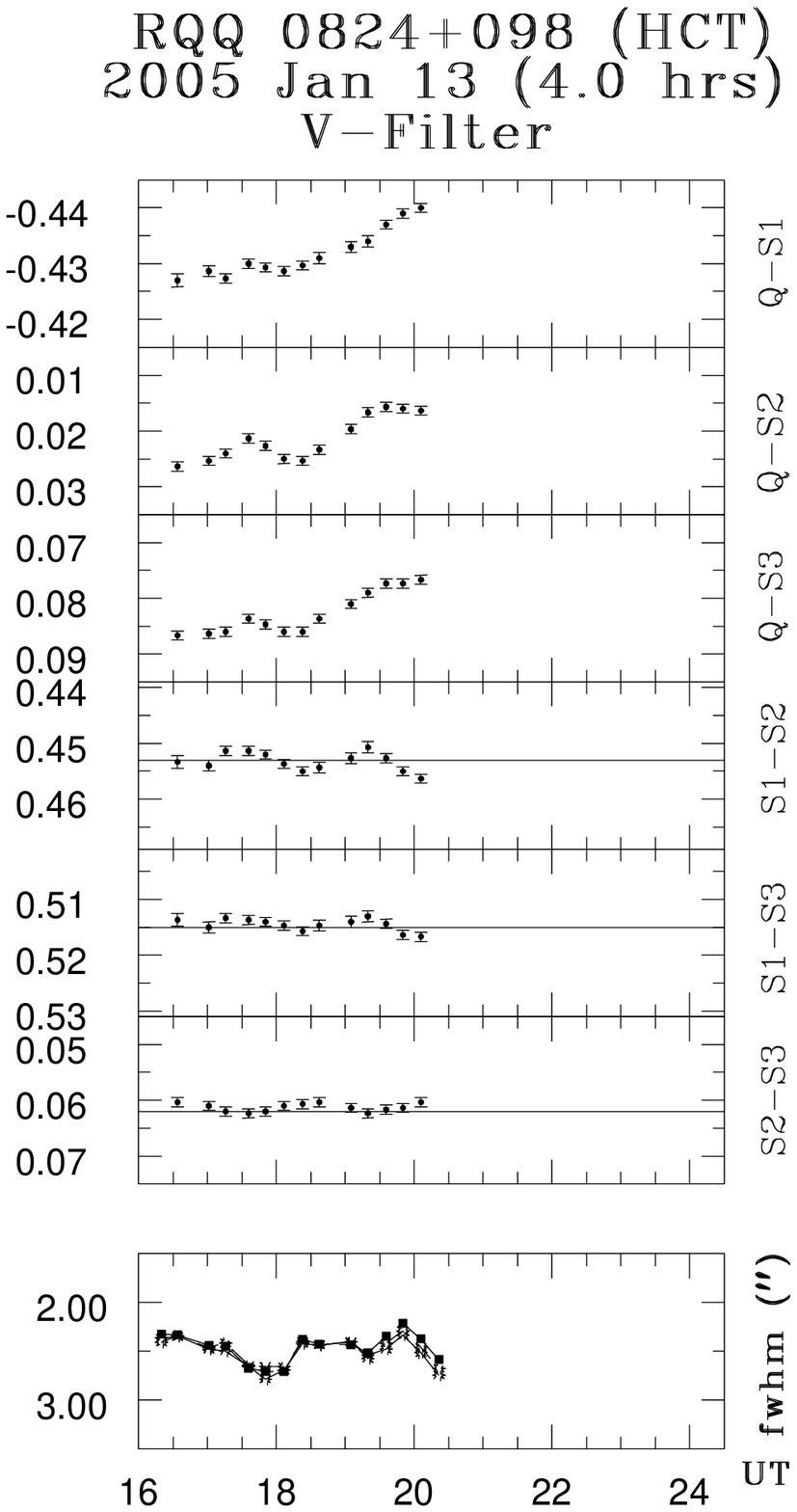}}
}
\caption{DLCs of radio-quiet quasars monitored
simultaneously with the ST and HCT telescopes.}
\label{fig:1}
\end{figure}
\newpage
\begin{figure}[h]
\hbox{
\hspace*{-15mm}{\includegraphics[height=10.0cm,width=07.0cm]{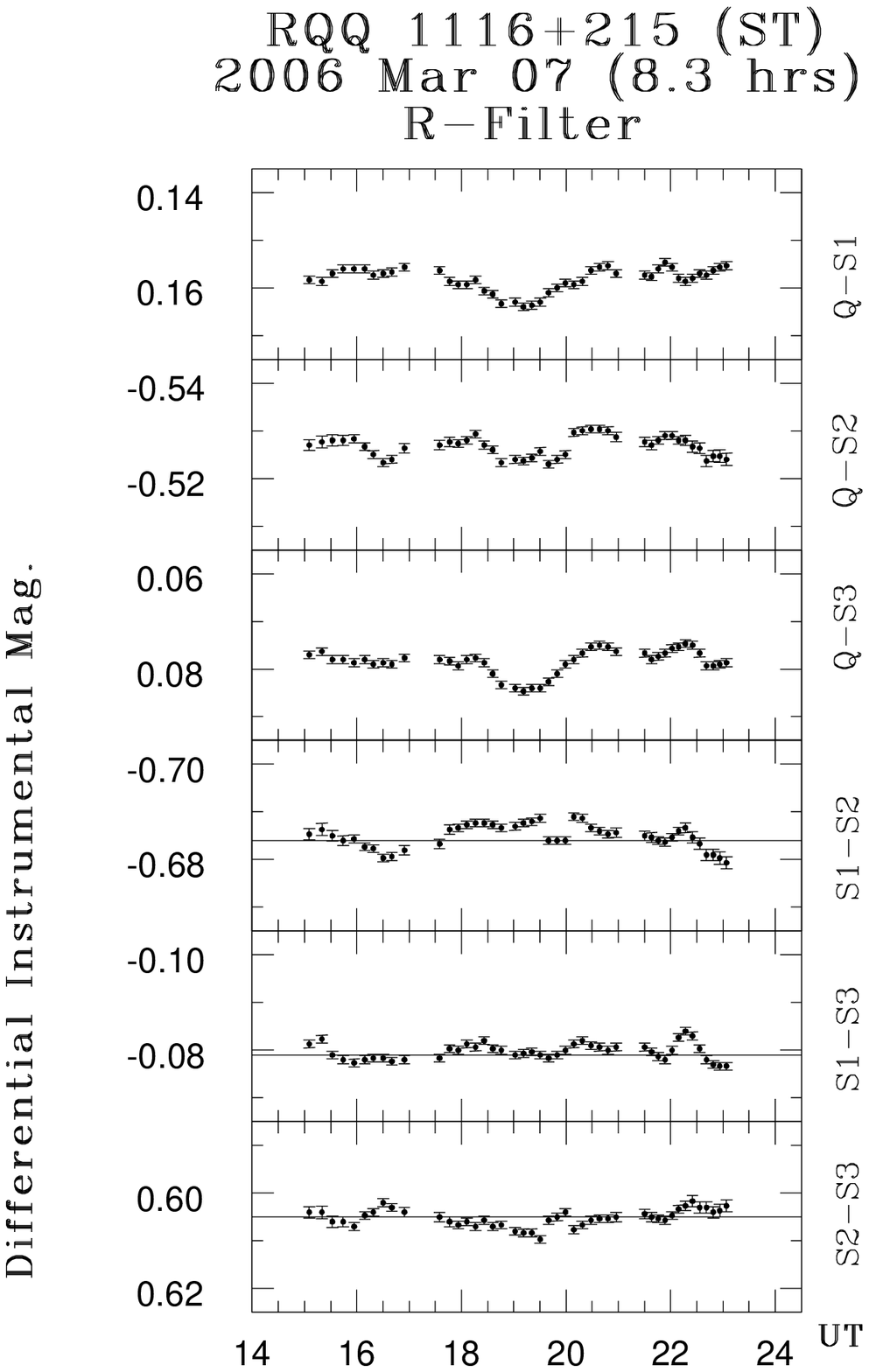}}
\hspace*{-30mm}{\includegraphics[height=10.0cm,width=07.0cm]{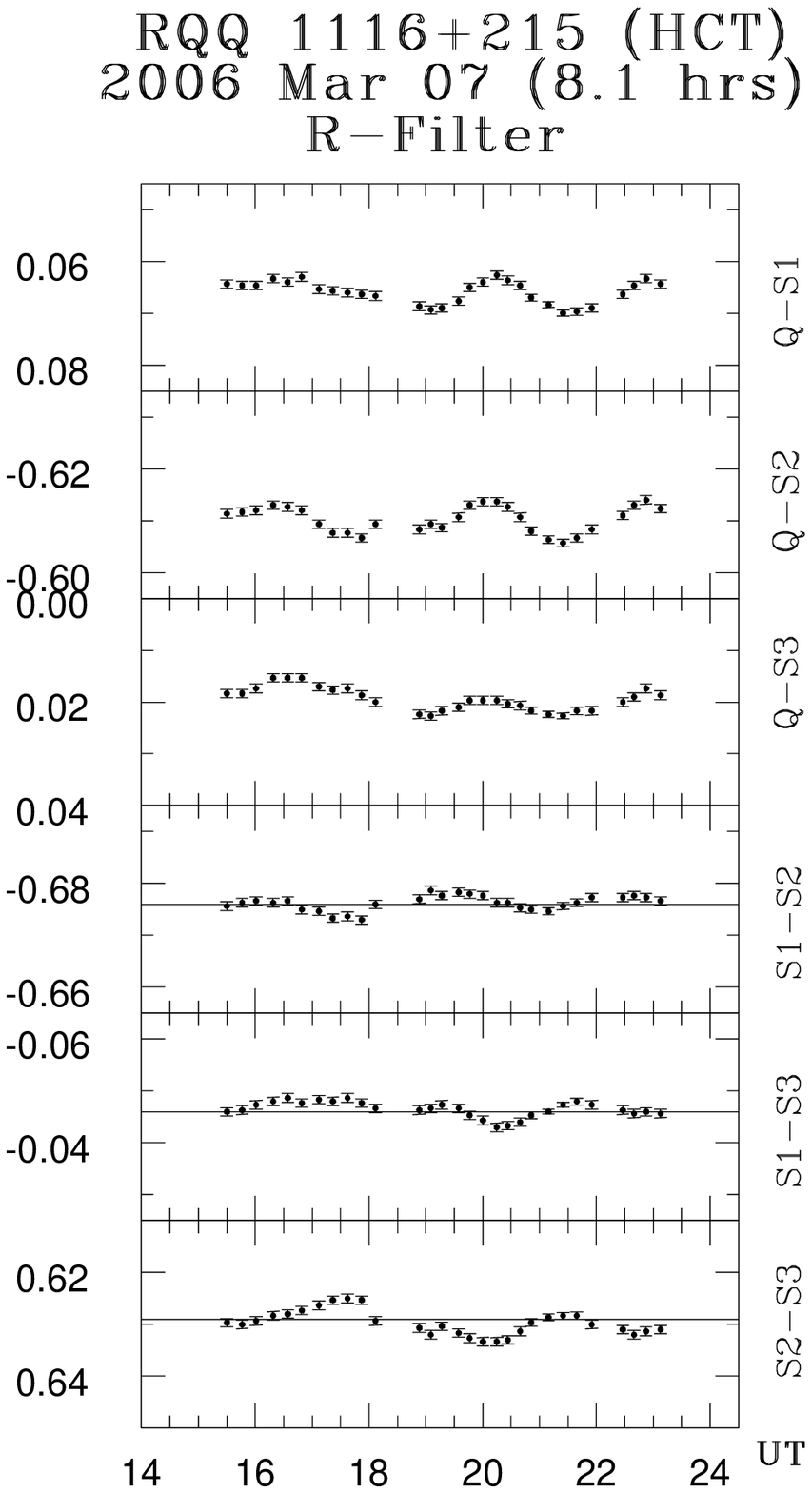}}
}
\begin{center}
{{\bf Figure~\ref{fig:1}}. \textit {continued}}
\end{center}
\end{figure}
\begin{figure}
\hbox{
\hspace*{-15mm}{\includegraphics[height=10.0cm,width=07.0cm]{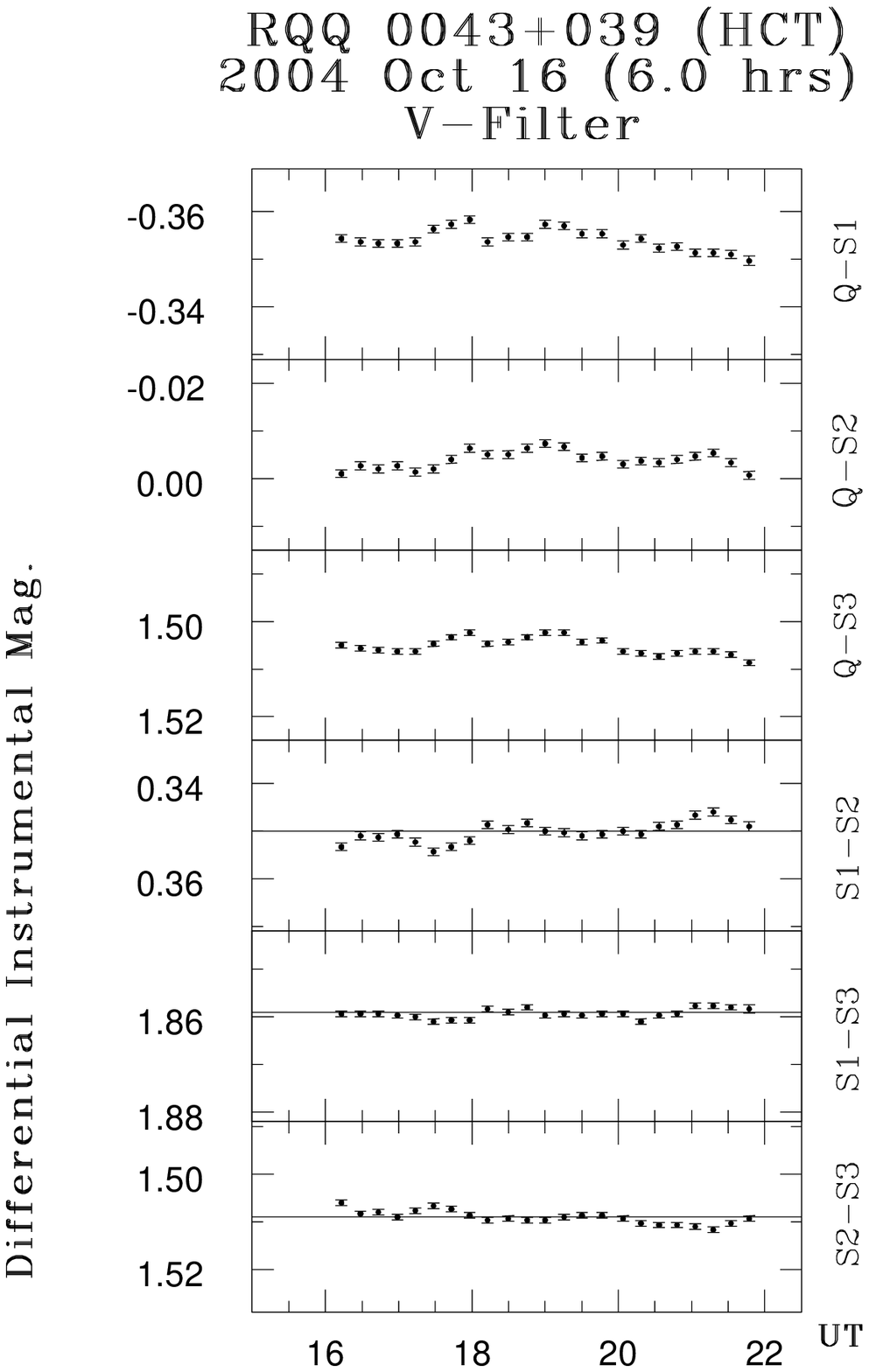}}
\hspace*{-30mm}{\includegraphics[height=10.0cm,width=07.0cm]{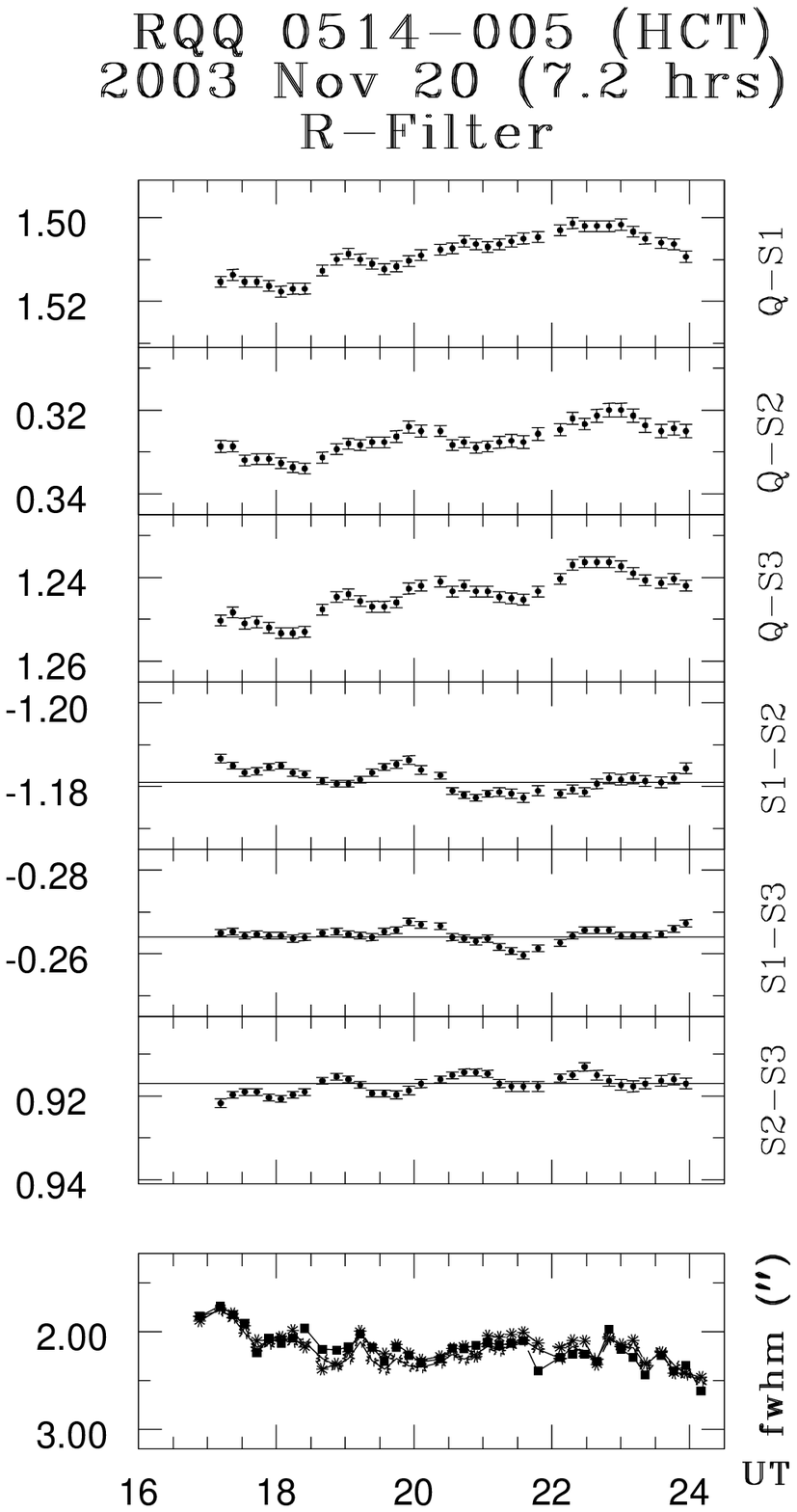}}
\hspace*{-30mm}{\includegraphics[height=10.0cm,width=07.0cm]{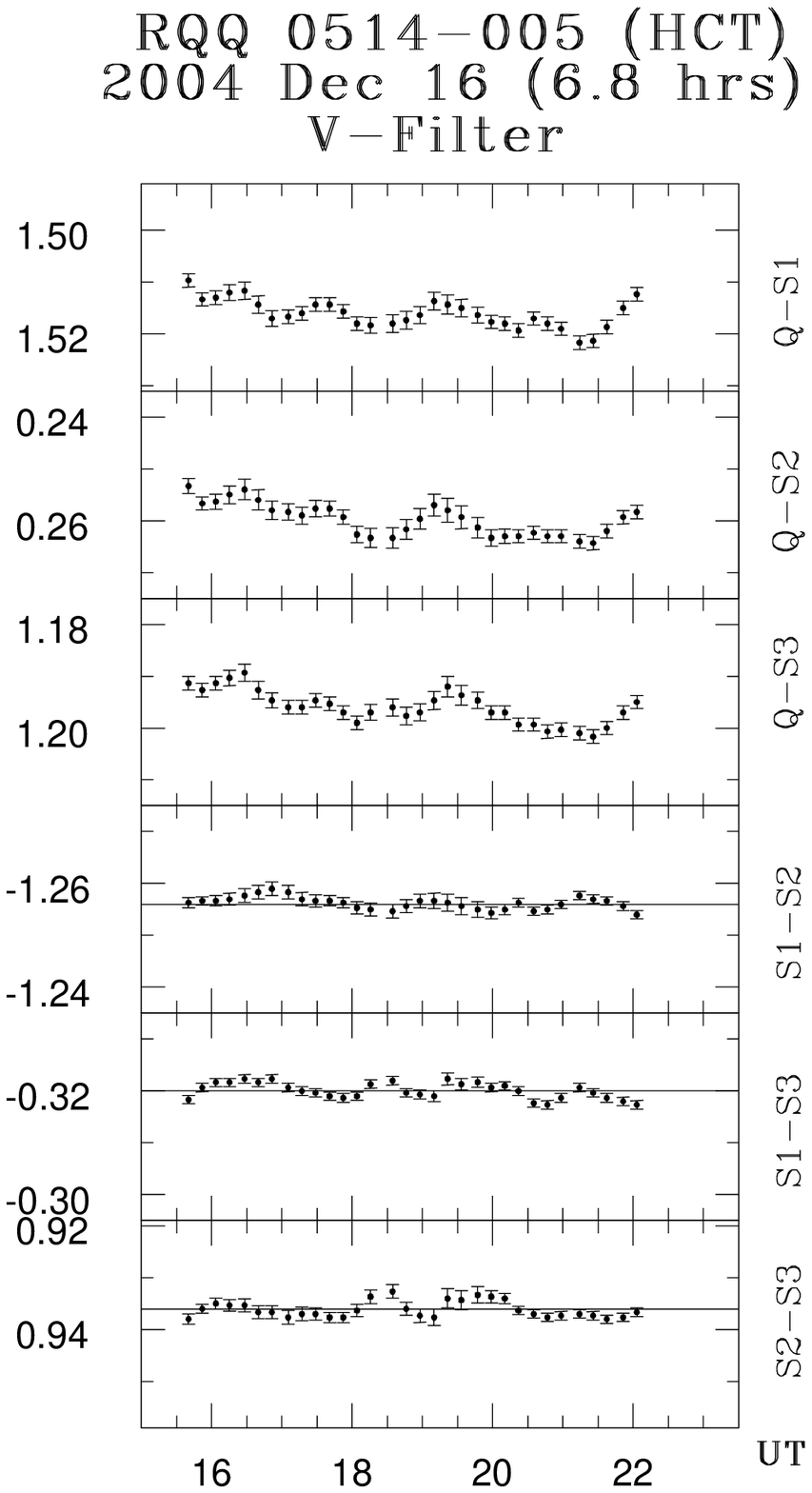}}
\hspace*{-30mm}{\includegraphics[height=10.0cm,width=07.0cm]{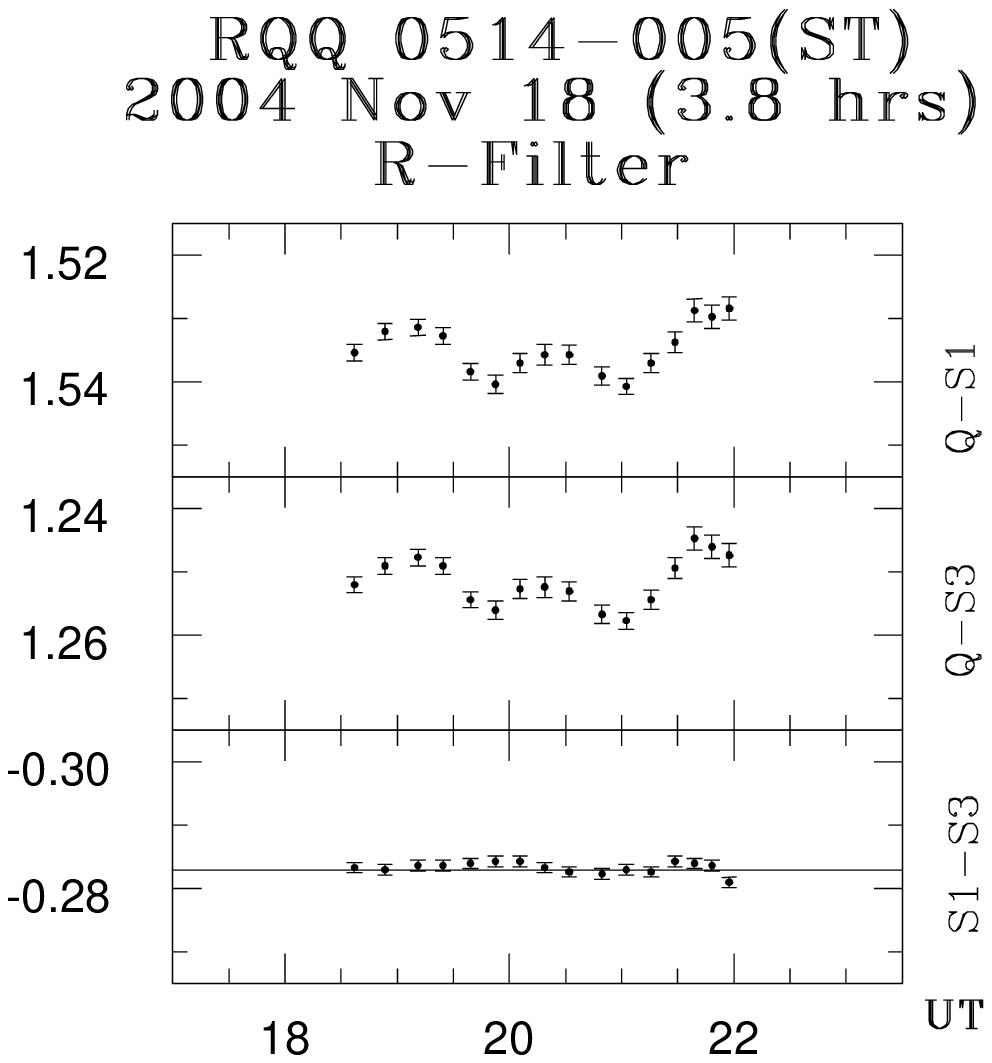}}
}
\hbox{
\hspace*{-15mm}{\includegraphics[height=10.0cm,width=07.0cm]{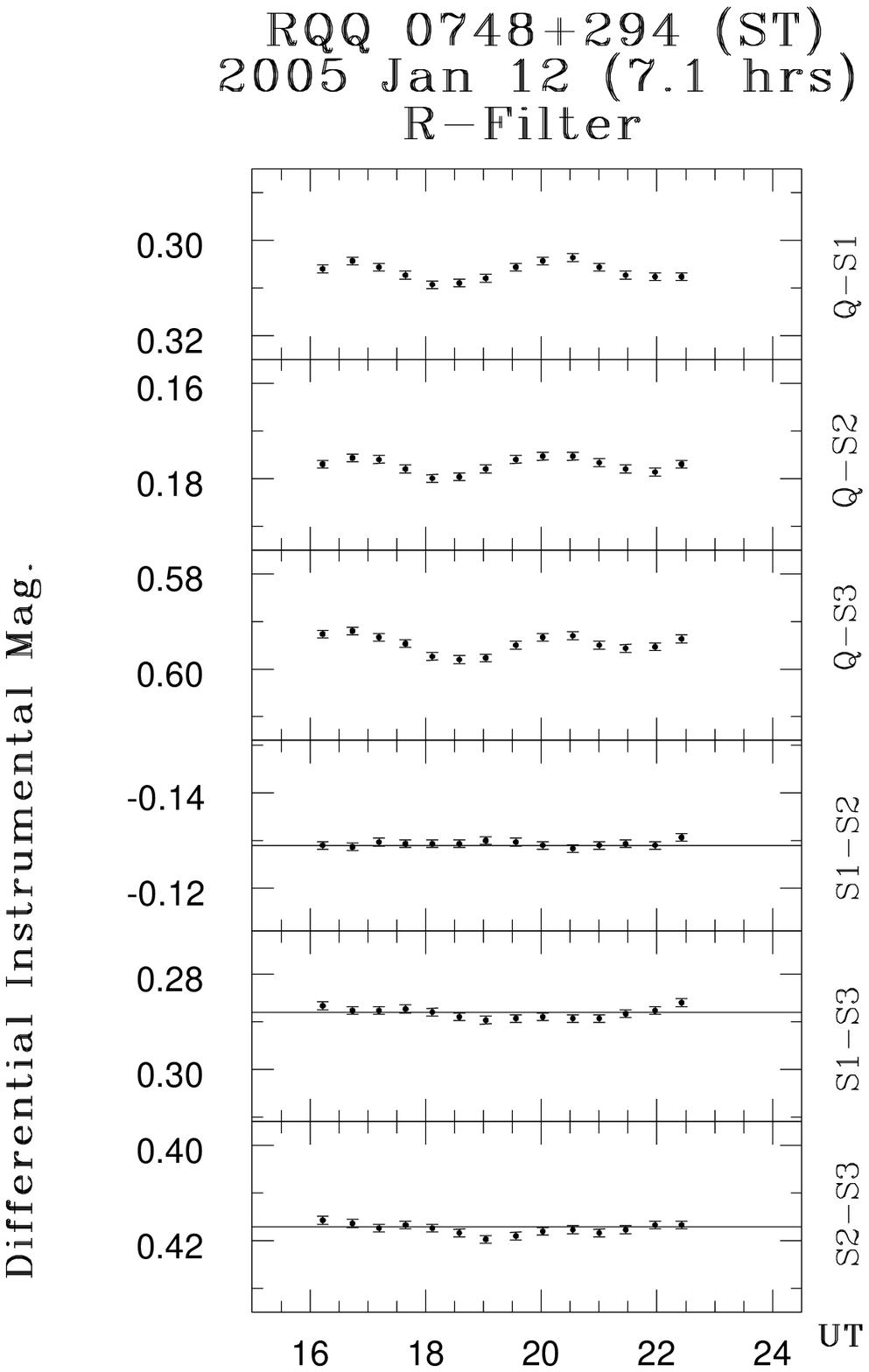}}
\hspace*{-30mm}{\includegraphics[height=10.0cm,width=07.0cm]{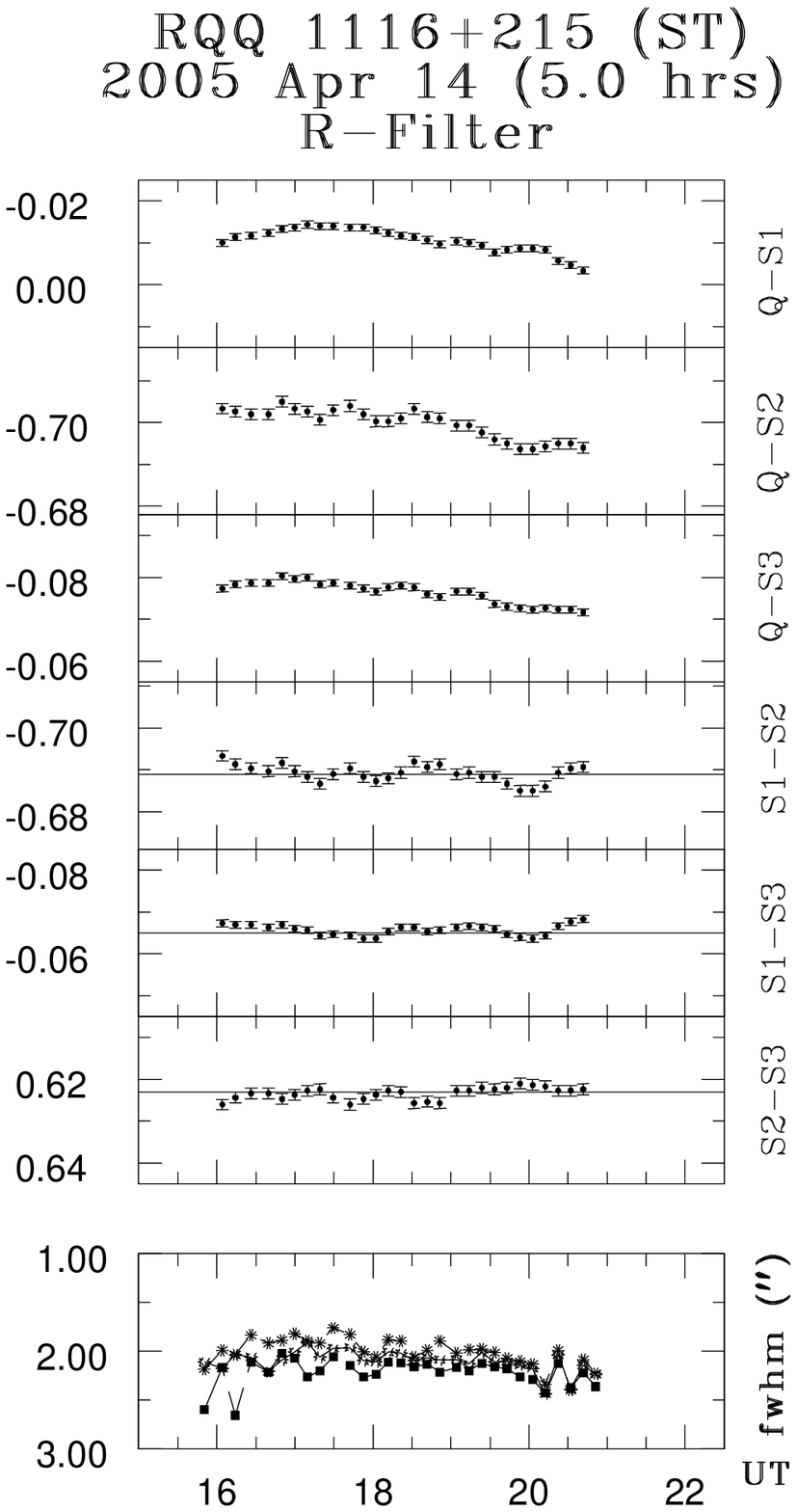}}
\hspace*{-30mm}{\includegraphics[height=10.0cm,width=07.0cm]{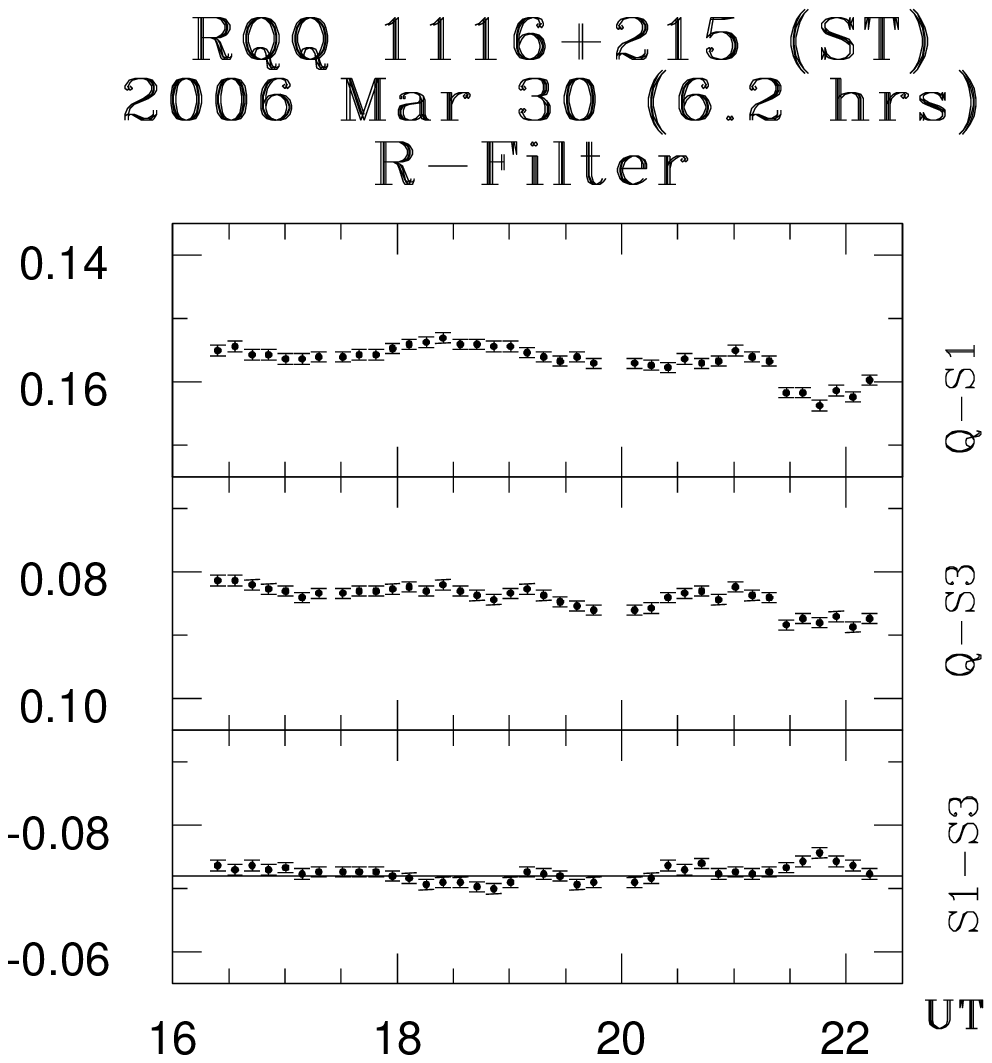}}
\hspace*{-30mm}{\includegraphics[height=10.0cm,width=07.0cm]{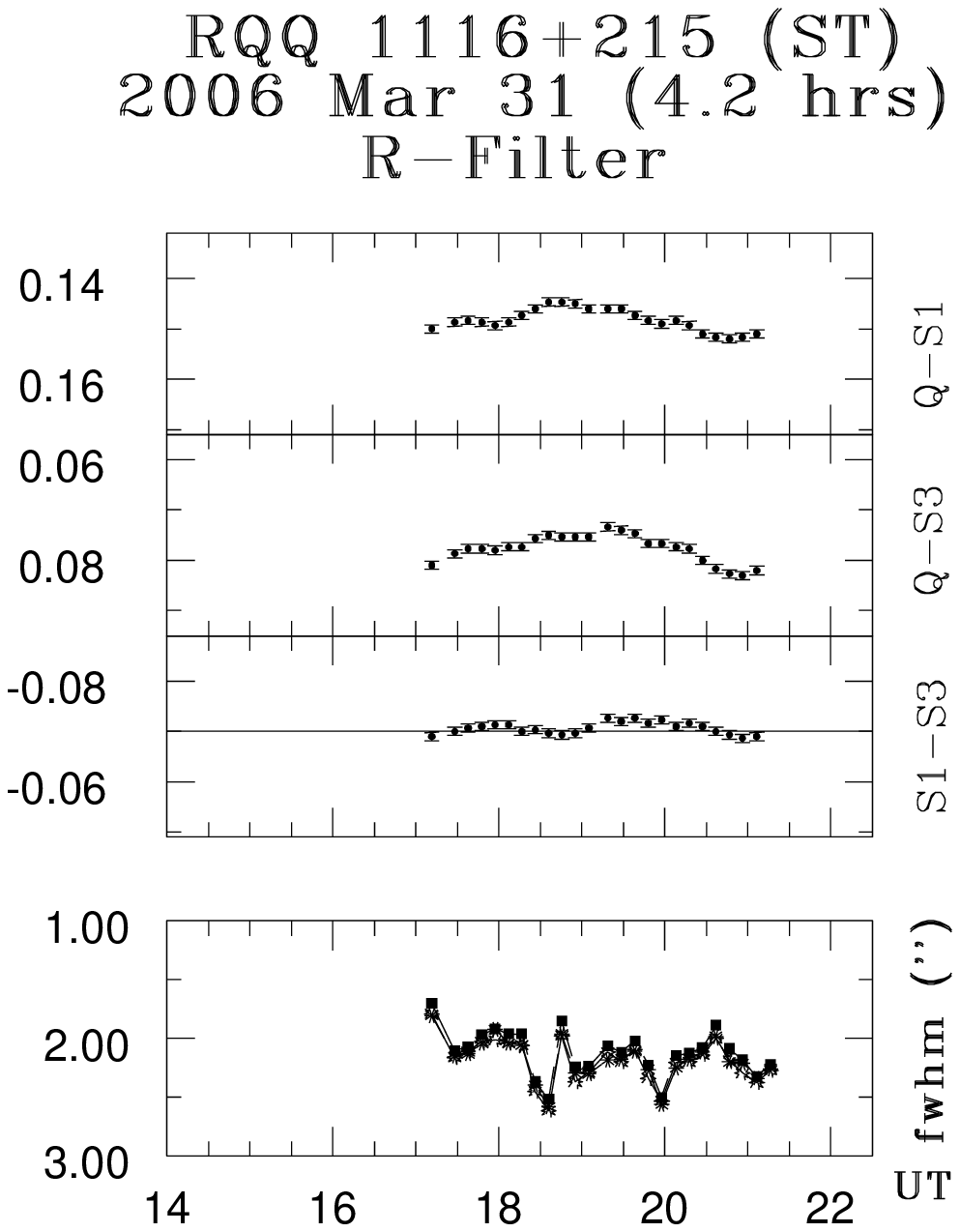}}
}
\caption{DLCs of radio-quiet quasars having
single station monitoring.}
\label{fig:2}
\end{figure}
\begin{figure}[h]
\hbox{
\hspace*{-15mm}{\includegraphics[height=10.0cm,width=07.0cm]{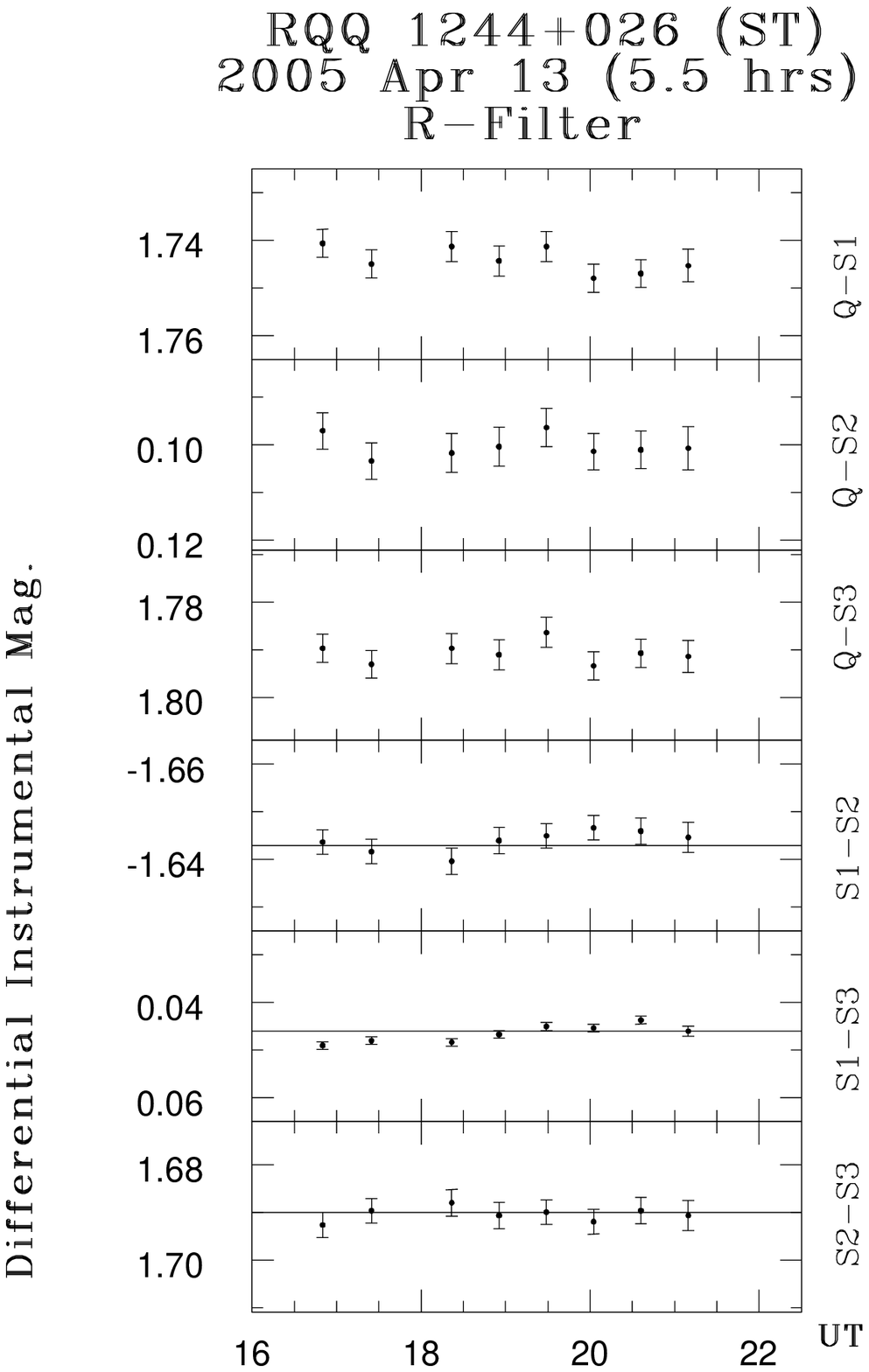}}
\hspace*{-30mm}{\includegraphics[height=10.0cm,width=07.0cm]{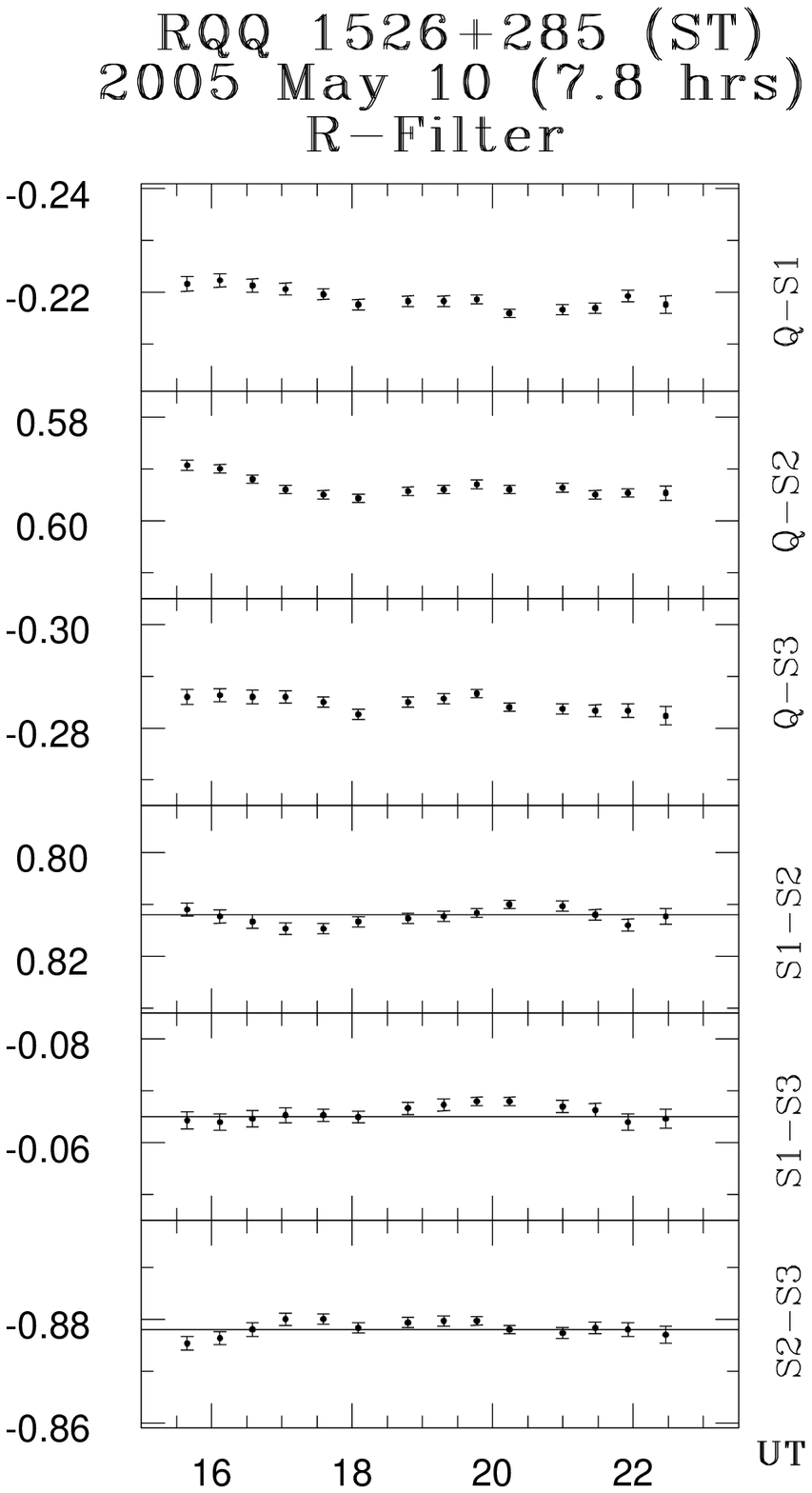}}
\hspace*{-30mm}{\includegraphics[height=10.0cm,width=07.0cm]{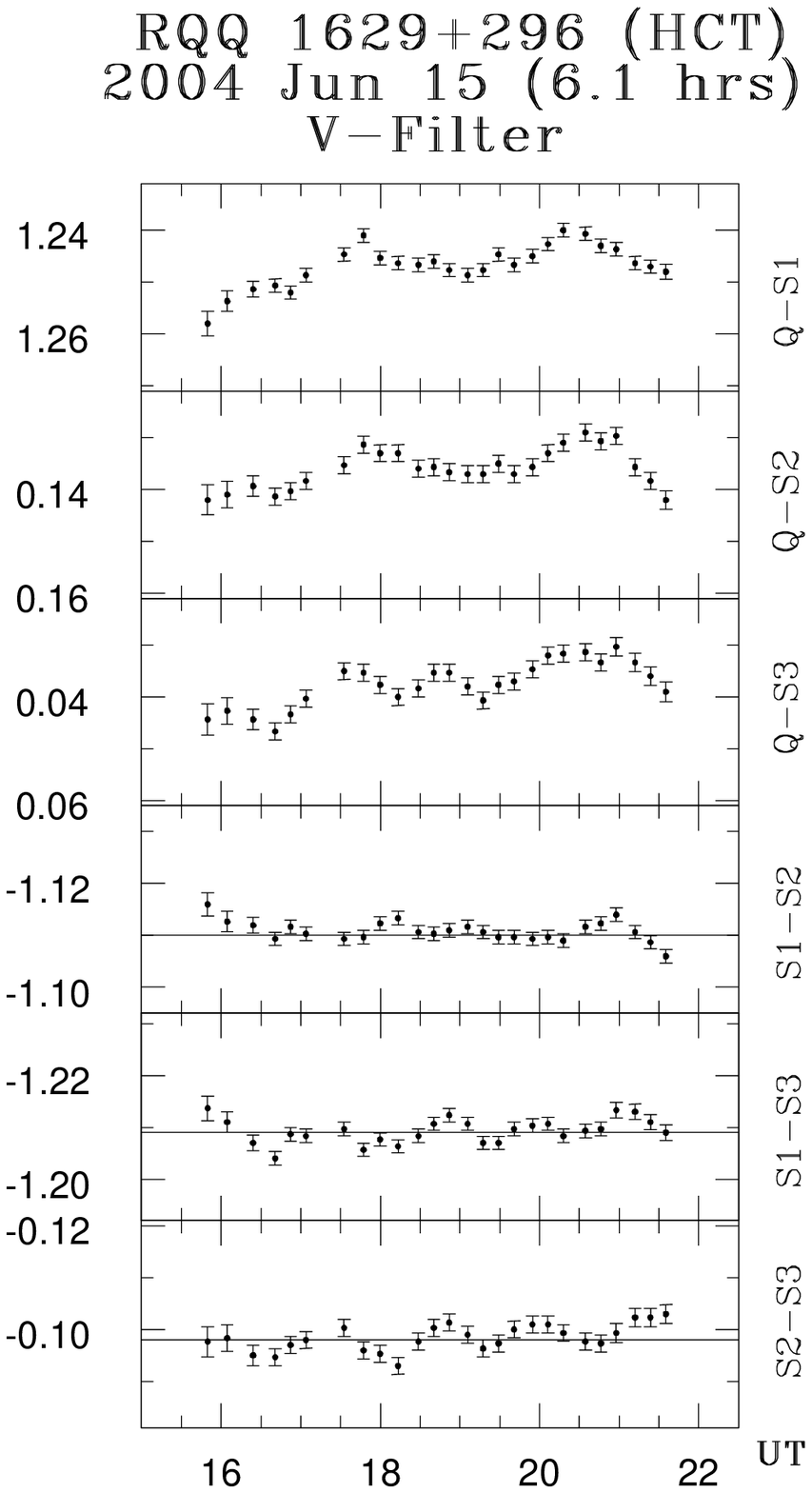}}
\hspace*{-30mm}{\includegraphics[height=10.0cm,width=07.0cm]{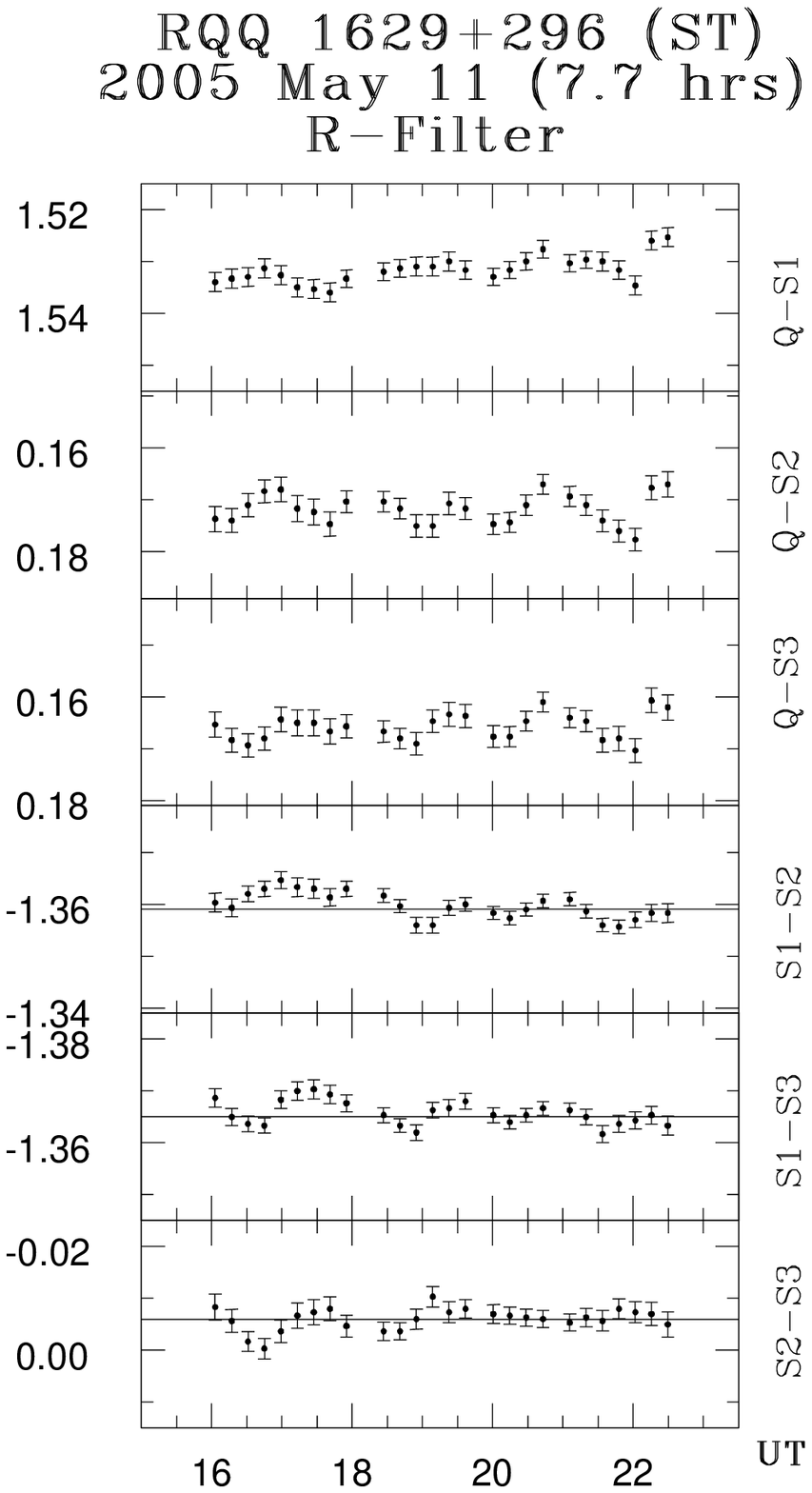}}
}
\hbox{
\hspace*{-15mm}{\includegraphics[height=10.0cm,width=07.0cm]{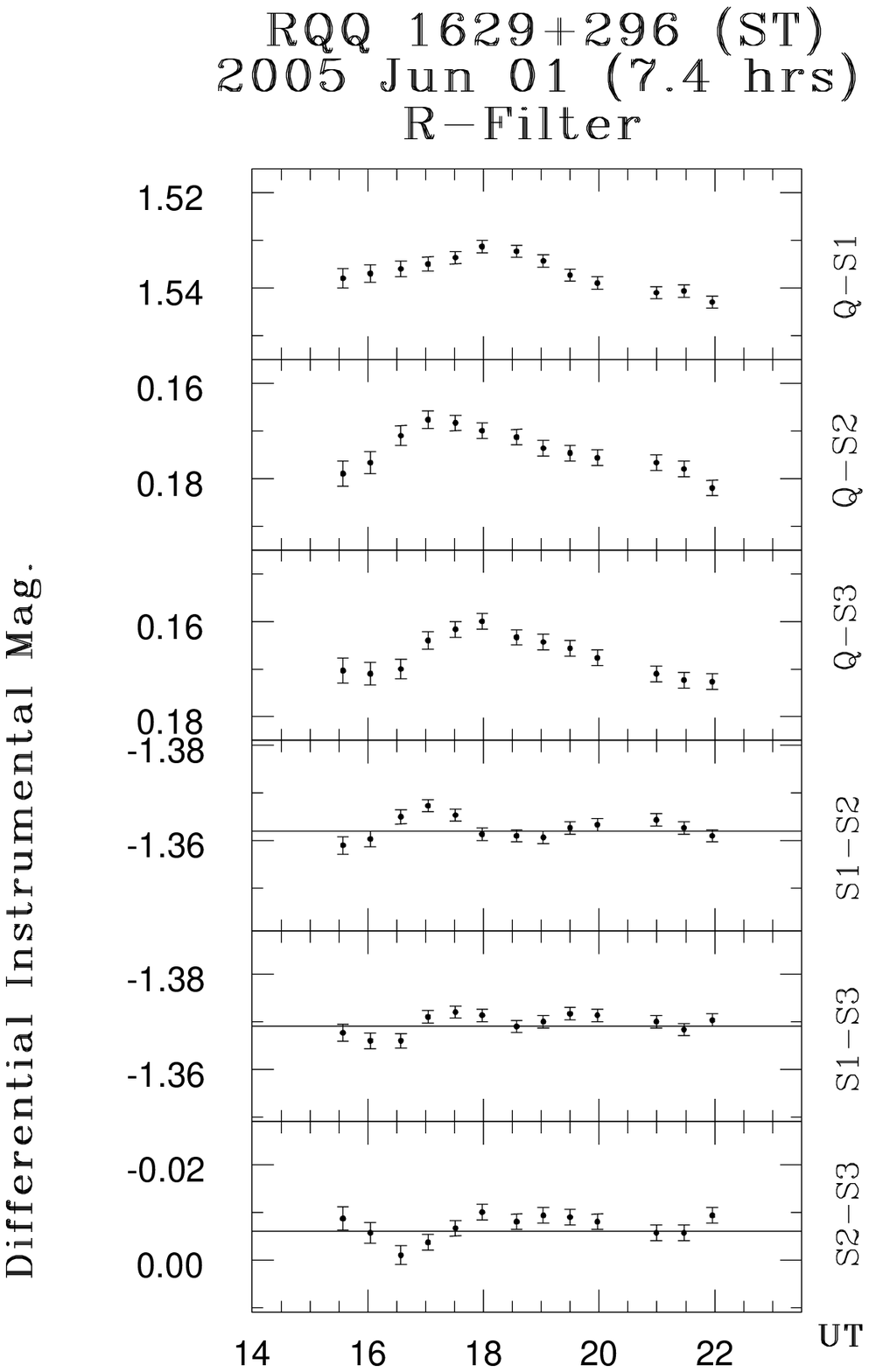}}
\hspace*{-30mm}{\includegraphics[height=10.0cm,width=07.0cm]{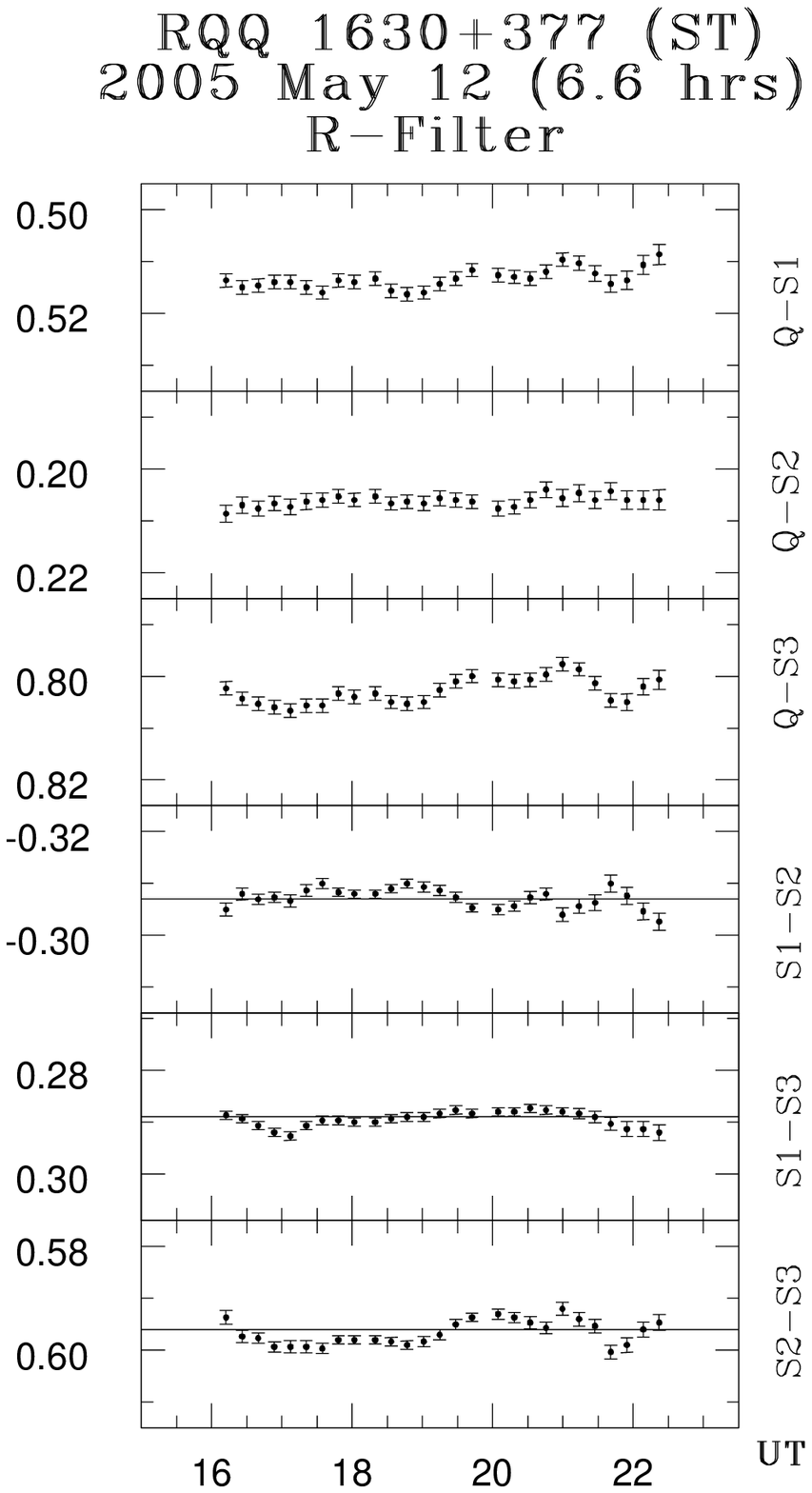}}
}
\begin{center}
{{\bf Figure~\ref{fig:2}}. \textit {continued}}
\end{center}
\end{figure}
\clearpage
\newpage
\label{lastpage}

\end{document}